\documentclass[pdflatex,sn-mathphys,Numbered]{sn-jnl}

\usepackage{graphicx}%
\usepackage{multirow}%
\usepackage{amsmath,amssymb,amsfonts}%
\usepackage{amsthm}%
\usepackage{mathrsfs}%
\usepackage[title]{appendix}%
\usepackage{xcolor}%
\usepackage{textcomp}%
\usepackage{manyfoot}%
\usepackage{booktabs}%
\usepackage{listings}%

\theoremstyle{thmstyleone}%
\theoremstyle{thmstyletwo}%

\theoremstyle{thmstylethree}%

\usepackage{iftex}

\usepackage[T2A,T1]{fontenc}
\DeclareTextFontCommand{\textcyr}{\fontencoding{T2A}\selectfont}

\usepackage{placeins}
\usepackage{flafter}
\graphicspath{{./img/}}

\let\OriginalTitlefont\Titlefont
\renewcommand{\Titlefont}{\OriginalTitlefont\centering}

\begin{document}

\title[Article Title]{Spectral dependence of Berry phase \\ and topological bound states in the continuum in M\"obius resonators}

\author[]{\fnm{Mikhail E.} \sur{Bochkarev}}\email{mikhail.bochkarev@metalab.ifmo.ru}
\equalcont{These authors contributed equally.}

\author[]{\fnm{Timur Z.} \sur{Seidov}}\email{tim.phys.math@gmail.com}
\equalcont{These authors contributed equally.}

\author[]{\fnm{Nikolay S.} \sur{Solodovchenko}}\email{n.solodovechenko@metalab.ifmo.ru}
\equalcont{These authors contributed equally.}

\author[]{\fnm{Kirill B.} \sur{Samusev}}\email{k.samusev@metalab.ifmo.ru}

\author*[]{\fnm{Mikhail F.} \sur{Limonov}}\email{m.limonov@metalab.ifmo.ru}

\affil[]{\orgdiv{School of Physics and Engineering}, \orgname{ITMO University}, \\ \orgaddress{\street{Kronverksky Pr. 49}, \city{St. Petersburg}, \postcode{197101}, \country{Russia}}}


\abstract{The M\"obius strip and the bulk M\"obius resonator are of great interest to specialists in optics, acoustics, and quantum physics. Despite the apparent simplicity of the object and numerous publications, the M\"obius resonator, like a magic box, still holds many surprises. Here, we demonstrate experimentally, theoretically and with a simple illustrative example that in a ``thin'' M\"obius resonator with an odd number of twists, the longitudinal wavenumber is determined by the additional phase shift $\pi$ (traditionally called the geometrical Berry phase) only if the total transverse index  of the photonic mode is odd. For an even index, the Berry phase is zero. We also discovered ``topological'' quasi-bound states in the continuum in M\"obius resonators with a square cross-section; they differ from ``ordinary'' ones by the exchange of azimuthal indices of the interfering modes in the anti-crossing region. Finally, we demonstrate an example of resonant topology by determining the dependence of the Berry phase on the M\"obius resonator geometry based on the frequency difference between the eigenmodes of planar ring resonators and those of M\"obius resonators. The discovered effects significantly expand our knowledge of the topology of single resonators and promise new practical applications.}


\maketitle


Curvature underlies the natural world around us and many artificially created objects that we describe using the fundamental laws of physics. The curvature determines many nontrivial physical properties \cite{Turner2010,Forgan2011}, including topologically induced magnetization patterning \cite{Kravchuk2012}, electronic properties of graphene \cite{Juan2011}, extrinsic curvature effects in liquid crystals \cite{Napoli2012}, shape-dependent metamaterials and metasurfaces \cite{Sarkar2019,Yu2011}. 

A classic object for studying the effects of curvature and topology is the non-orientable, one-sided surface, independently described by Johann Listing and August M\"obius in 1858. Called the M\"obius strip (M\"oS), this object is used in fine art and architecture and is actively studied in physics and chemistry, both theoretically and experimentally \cite{Tanda2002,Starostin2007,Ballon2008,Fomin2012,Li2012,Pylypovskyi2015,Yin2017,Kreismann2018,Nishiguchi2018,Lin2022}. A M\"oS is a 2D single-sided surface formed by connecting the two ends of a rectangular strip, twisted into an odd number of twists \cite{Tanda2002,Starostin2007}. As is well known, a M\"obius ring (M\"oR) of finite thickness has two sides, representing two M\"oSs, rather than four sides like a regular planar ring resonator (RR). 

Here, we pay special attention to the additional phase of photonic eigenmodes in non-trivially twisted light guides, which was first described in \cite{Pancharatnam1956,Berry1984,Berry1987} and subsequently called the ``Berry phase'' (BPh). It has been shown that during the propagation of a polarized light beam, the phase change is determined by two contributions: the well-known dynamical phase, which depends on the propagation time, and the BPh, which depends on the geometry of the light guide. The BPh has been studied in a number of papers \cite{Cohen2019,Tomita1986,Yale2016,Khanikaev2017,Song2020,Chen2023}, including in a M\"oR by varying the height-to-width ratio \cite{Wang2022}. When linearly polarized light enters a thin M\"oR, the electric field is forced to remain parallel to the plane of the twisted structure due to predominantal in-plane confinement and therefore the polarization orientation continuously changes, resulting in a $180^\circ$ rotation of the field vector E in the case of one twist. Thus, a single twist adds a geometric BPh $\pi$, so the condition for constructive interference in a thin M\"oR is satisfied for a half-integer number of waves along the M\"oR's length. However, as the height-to-width ratio of the M\"oR increases, the polarization of light changes from linear to elliptical, and in a M\"oR with a square cross-section, the total BPh becomes zero, which was demonstrated experimentally \cite{Wang2022}. 

The M\"oR has a long history of extensive research, but interest in it has not waned, and the number of publications has not decreased in recent years \cite{Chen2023,Wang2022,Paterson2026,Sun2024,Thonikkadavan2026,Dai2026}. However, despite numerous publications, in this field still there are areas to explore. In this paper we significantly extend the generally accepted understanding of the photonic properties of the M\"oR and Berry's geometrical phase. New results were obtained by applying geometric holonomy to photonic eigenmodes with different symmetries, as well as through numerical and experimental studies of the extinction spectra of M\"oRs. In this paper, we demonstrate that the BPh in a thin single-twist M\"oR varies along the spectrum and is either the well-known value of $\pi$ or zero, depending on the sum of the radial $(r)$ and axial $(z)$ indices of the longitudinal modes. 

The second aspect we are interested in is quasi-bound states in the continuum (q-BIC) \cite{Hsu2016} in M\"oRs. We have previously studied q-BICs in topologically trivial dielectric structures such as cylinders \cite{Rybin2017,Bogdanov2019}, RRs \cite{Solodovchenko2021}, split RRs, and cuboids \cite{Solodovchenko2025}, the dependence of q-BICs on material losses was also studied \cite{Bochek2022}. Here, we are interested in the occurrence of q-BICs in M\"oRs with different cross-sections. We fix the size of one side (M\"oS$_\mathrm{W}$), hereinafter called the \textit{width} ($\mathrm{W}_\mathrm{f}$), and vary the size of the other side (M\"oS$_\mathrm{H}$), called the \textit{height} ($\mathrm{H}$), starting with a thin M\"oR ($\mathrm{W}_\mathrm{f}>\mathrm{H}$), passing through a square cross-section ($\mathrm{W}_\mathrm{f}=\mathrm{H}$), and ending with a tall rectangular M\"oR ($\mathrm{W}_\mathrm{f}<\mathrm{H}$). 

In the M\"oRs extinction spectra, we identified two different types of q-BICs. First, ``ordinary'' q-BICs in rectangular-cross-section M\"oRs, similar to those observed in RR \cite{Solodovchenko2021,Solodovchenko2025},where each photonic mode with specific indices ($r,z,m$) preserve its dispersion dependence in parametric space (frequency versus $\mathrm{H}/\mathrm{W}_\mathrm{f}$) despite numerous anti-crossings with other modes. Secondly, the q-BICs, which we termed ``topological'' because they arise from the unique nature of the field distribution in a M\"oR with a square cross-section, leading to an exchange of azimuthal indices ($m_{\mathrm{radial}} \! \! \leftrightarrow \! \! m_{\mathrm{axial}}$) between interfering radial and axial modes following their anti-crossing.

Finally, by combining the dependencies of the M\"oR and RR photonic modes, we demonstrate the dependence of the BPh on the M\"oR geometry: it remains constant at $\pi$ over a wide range of rectangular cross-sections, whereas in the narrow region around the square cross-section, it varies, passing through zero at $\mathrm{W}_\mathrm{f}=\mathrm{H}$.

\FloatBarrier
\section*{Results}\label{sec2}

\FloatBarrier
\subsection*{Spectral dependence of Berry phase: Experiment}\label{subsec2.1}

%
\begin{figure}[!htbp]
   \centering
   \includegraphics[width=\linewidth]{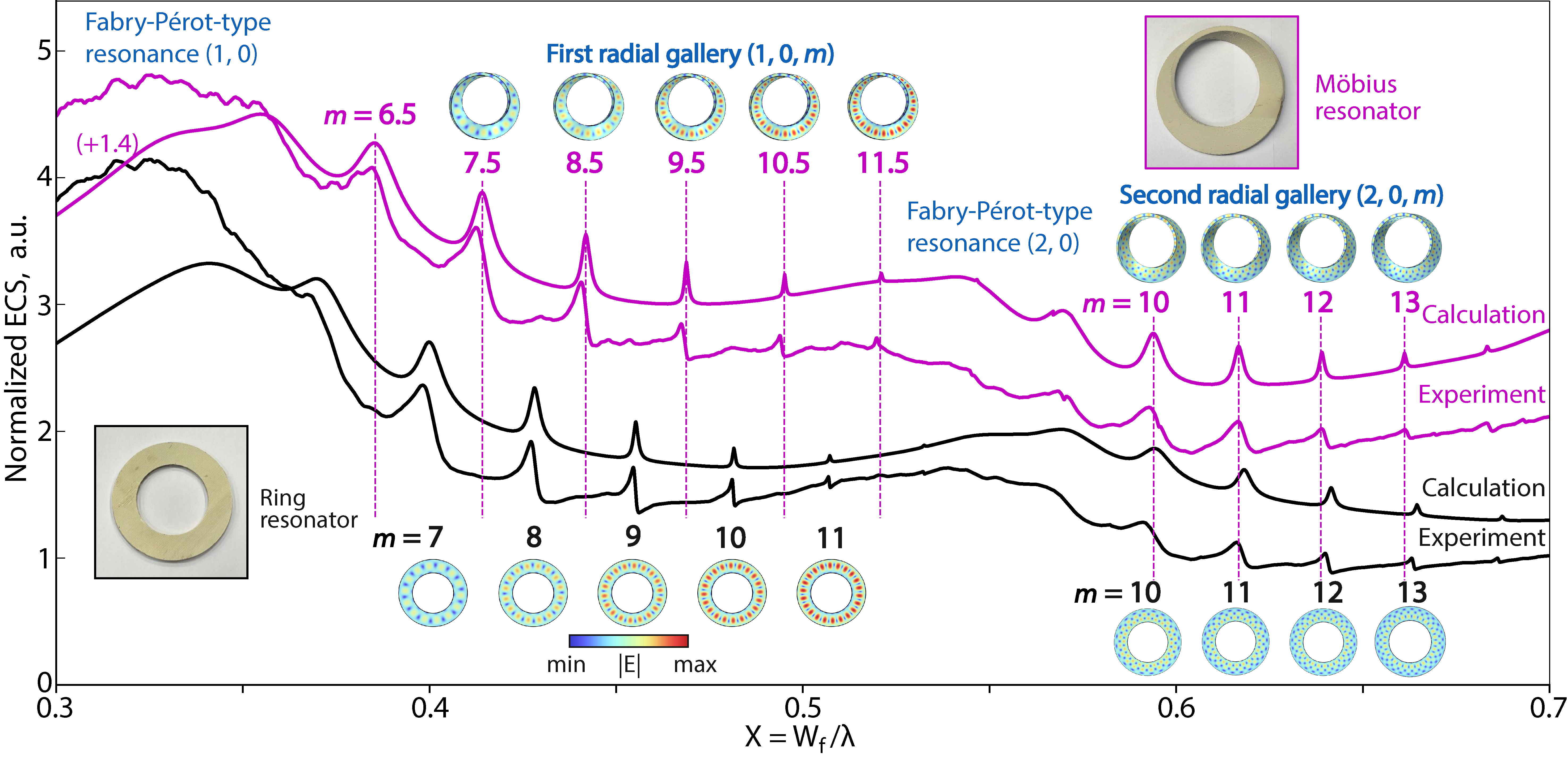}
   \caption
{\textbf{Extinction spectra of planar RR and M\"oR with one twist: calculations and experiment.} Experimental and calculated extinction spectra for a planar RR (black) and a M\"oR with one twist (red) in the spectral region of the first ($r=1, z=0, m$) and second ($r=2, z=0, m$) radial photonic galleries. The difference between the symmetric calculated and the asymmetric experimental lines is attributed to Fano resonance \cite{Limonov2017} between the incident and scattered waves. The M\"oR spectra are shifted vertically by $1.4$ a.u. The calculated fields $|\mathbf{E}|$ on the surface of the resonators and azimuthal indices $m$ are given above and below the corresponding resonance lines in the spectra. The calculated fields are shown on an enlarged scale in Fig.~2s in the Supplementary, Sec.~2. The dimensionless frequency $\mathrm{X} = \mathrm{W}_\mathrm{f} / \lambda$. Insets show photographs of experimental samples. Both resonators have the same rectangular cross-section, with $\mathrm{W}_\mathrm{f}>\mathrm{H}$ ($20\mathrm{x}4 \, \mathrm{mm}^2$), the same length along the midline, and the same permittivity $\epsilon_{res}=8.23$, $\tan \delta_{res} =1.1 \cdot 10^{-4}$, $\epsilon_{env}=1$, see Methods and \cite{Filament}.
}
\label{fig:fig01}
\end{figure}
%

Figure~\ref{fig:fig01} presents the calculated and experimental extinction spectra of RR and M\"oRs with one twist. The samples were fabricated by 3D printing, see Methods. Methods for calculating spectra and distribution of resonance fields are presented in Methods and Supplementary, Sec.2, the experimental technique is described in Methods. We have previously demonstrated that the scattering spectra of dielectric RRs with a rectangular cross-section are divided into radial and axial galleries \cite{Solodovchenko2022}. Each photonic gallery begins with broadband Fabry-P\'erot-type resonances between the side walls (radial indices $r$) or between the upper and lower walls (axial indices $z$), and continues with equidistant narrowband longitudinal resonances with an azimuthal index $m$. 

The fundamental difference between M\"oR and RR is that RR is formed by four sides, and the pairs of sides are divided into two lateral and top-bottom, therefore the concepts of radial and axial photonic modes have a clear definition and do not change with any changes in the geometry of RR. M\"oR has only two sides (two M\"obius strips), and therefore the concepts of radial and axial modes lose physical meaning. Nevertheless, our results demonstrate that, in the case of thin M\"oR ($\mathrm{W}_\mathrm{f}=5\mathrm{H}$), separating the observed lines into radial modes (corresponding to a wide side, M\"oS$_\mathrm{W}$) and axial modes (a narrow side, M\"oS$_\mathrm{H}$) simplify the interpretation of the spectra. In Fig.~\ref{fig:fig01}, the low-frequency extinction spectrum contains two radial galleries with indices $(r=1, z=0, m)$ and $(r=2, z=0, m)$, while the axial galleries $(z=1, 2 \dots)$ at $\mathrm{W}_\mathrm{f}=5\mathrm{H}$ are located in the higher-frequency region of the spectrum. This interpretation is based on calculations of frequencies and electromagnetic fields of resonant modes, which allow the indices $(r,z,m)$ to be unambiguously determined, Supplementary, Sec.~2.

Our interest is in the relative spectral position of the azimuthal modes in the two galleries in the spectra of the single-twist M\"oRs with respect to the azimuthal modes of the RR. The crucial result, shown in Fig.~\ref{fig:fig01}, is that the azimuthal modes of the single-twist M\"oR behave differently depending on the gallery, that is, on the indices $(r,z)$. We demonstrate that the azimuthal modes of the first gallery $(r=1, z=0, m)$ are shifted by half the intermode period relative to the entire sequence of azimuthal modes of the RR, and, unexpectedly, the modes of the second gallery $(r=2, z=0, m)$ are not shifted at all relative to the RR modes. Calculations of electromagnetic fields in resonators, which allow us to determine integer and half-integer longitudinal indices $m$ in M\"oRs and RRs, fully confirm the appearance of half-integer indices m in the first gallery and integer indices $m$ in the second gallery of the M\"oR with one twist, Supplementary, Fig.~2s. Hereafter, we will refer to photonic galleries with integer longitudinal indices m as ``trivial'' and to those with non-integer indices m as ``topological''.  

The effect of the modes shift of the first (topological) gallery of M\"oR with one twist corresponds to the generally accepted picture of photonic effects in thin $(\mathrm{W}_\mathrm{f}>\mathrm{H})$ M\"oR \cite{Wang2022}. In a thin M\"oR, during propagation of a linearly polarized wave, the vector $\mathbf{E}$ is oriented strictly along the long wall. When the resonator is twisted, the electric field rotates in space along with the resonator itself. In the case of a single twist ($180^\circ$ rotation), after traveling the entire path, the vector $\mathbf{E}$ is also rotated $180^\circ$ relative to the initial orientation, which corresponds to an additional geometric BPh of $\pi$ \cite{Wang2022}. This means that a mode that is resonant in a RR with a wavelength $m \lambda _\mathrm{RR} = L_\mathrm{RR}$ and is determined only by the dynamic phase shift along the RR of $2 \pi m$ will not be resonant in a M\"oR due to the additional phase of $\pi$. Accordingly, the resonant modes in a thin M\"oR with one twist will be determined by the total dynamic and geometric phase condition $2 \pi m + \pi=2 \pi \cdot (m+1/2)$ and all resonant wavelengths will be shifted by half the intermode distance relative to a RR. Thus, the main mystery is the integer indices $m$ in the second (trivial) gallery of M\"oR with one twist, which suggests the presence of an additional mechanism not described by the Berry geometric phase alone.

\FloatBarrier
\subsection*{Spectral dependence of Berry phase: Theory}\label{subsec2.2}

We model the M\"obius resonator in Fig.~\ref{fig:fig01} as a dielectric waveguide closed along a circle of radius $R$. Its rectangular cross-section turns uniformly through $N$ twists over one circuit. The homogeneous, isotropic MöR of refractive index $n$ is in a low-index environment. We neglect material dispersion and absorption and assume negligible radiative loss over one circuit. Using the Frenet--Serret frame and the induced metric~\cite{Yabe1984,Takagi1992,Fomin2025}, we write the source-free Maxwell equations in coordinates that follow the cross-section. For transverse dimensions small compared with $R$, we follow an isolated bound mode of the corresponding straight guide adiabatically around the resonator. Its transverse profile satisfies the dielectric interface conditions at the resonator boundary (Methods and Sec.~4 of the Supplementary).

After $N$ twists, the transverse profile of a mode in gallery $(r,z)$ returns with a factor $(-1)^{N(r+z)}$, giving the BPh $\gamma_{rz}=N\pi(r+z)\pmod{2\pi}$. The longitudinal phase factor $e^{2\pi i m}$ must compensate for this sign for the field to be single-valued, where $m=qR$ and $q$ is the longitudinal propagation constant. For one twist, the $(r=1,z=0)$ gallery therefore has a BPh of $\pi$ and half-integer $m$, while the $(r=2,z=0)$  gallery has zero BPh and integer $m$. For nearly linear modal dispersion over one RR resonance spacing, the $(1,0,m)$ resonances lie approximately halfway between adjacent RR resonances of the same gallery, while the $(2,0,m)$ resonances have no leading geometric shift, in agreement with Fig.~\ref{fig:fig01}. 

Next, we investigate the interference of photonic galleries with transverse indices $(r = 1, z = 0)$ and $(r = 0, z = 2)$, Fig.~\ref{fig:fig02}. Once again, numerical calculations show that gallery $(r = 1, z = 0, m)$ is topological, with half-integer indices m, whereas gallery $(r = 0, z = 2, m)$ is trivial, with integer indices m.

\FloatBarrier
\subsection*{``Ordinary'' q-BICs in M\"oRs: anti-crossing of topological and trivial galleries }\label{subsec2.3}

%
\begin{figure}[!htbp]
   \centering
   \includegraphics[width=0.90\linewidth]{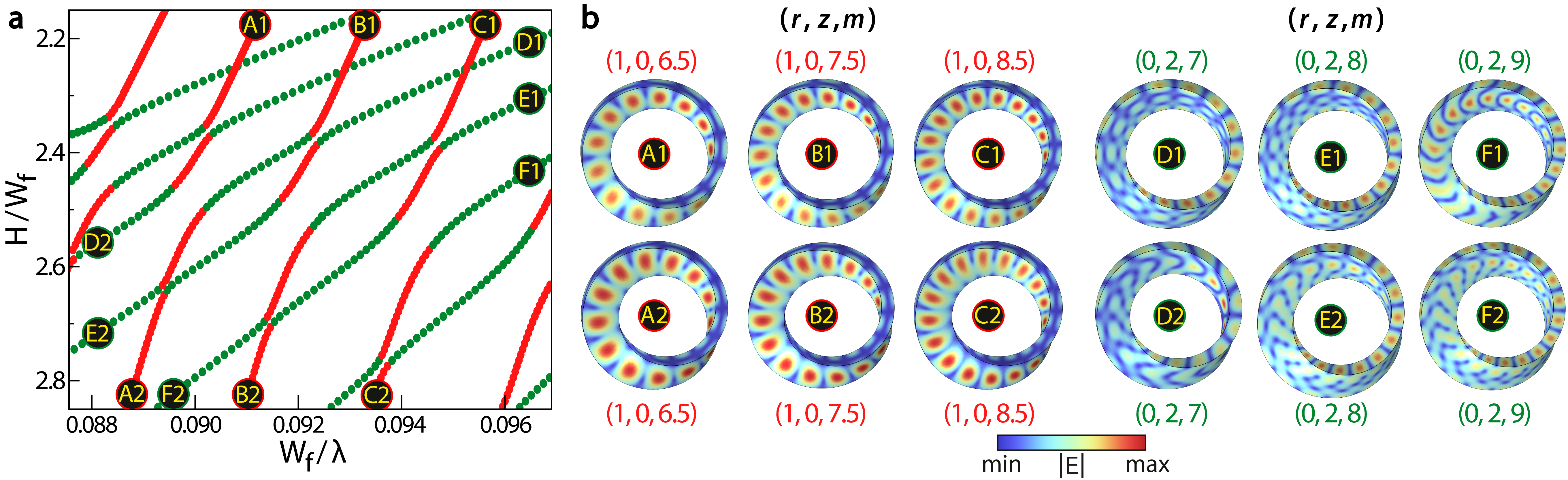}
   \caption
{\textbf{Preservation of the dispersion dependence of the topological radial $(1, 0, m)$ and trivial axial $(0, 2, m)$ modes in the region of ``ordinary'' cascades of q-BICs of M\"oRs.} \textbf{a}, Dispersion of the real parts of the topological radial (red, with a half-integer $m$) and trivial axial (green, with an integer $m$) modes in the region of multiple anti-crossings caused by q-BICs. As the height $H$ changes, the radial and axial gallery modes repeatedly anti-intersect, leading to multiple q-BIC cascades. The six modes are numbered: three radial (A, B, C) and three axial (D, E, F). \textbf{b}, Distribution of fields for six selected modes $(r, z, m)$ at the initial and final points of the parametric space region selected in panel ({a}). $\epsilon_{res}=40$, $\tan \delta_{res} =10^{-4}$, $\epsilon_{env}=1$. 
}
\label{fig:fig02}
\end{figure}
%

A striking example of cascades of q-BICs was discovered in the case of interference between radial and axial galleries of photonic azimuthal modes in dielectric resonators with a rectangular cross-section (RRs, split RRs, and cuboids) \cite{Solodovchenko2025}. When only one of the transverse dimensions of the resonator changes, only one type of gallery shifts significantly in the spectrum, demonstrating anti-crossing with the other and forming cascades of equidistant q-BICs described by the Friedrich-Wintgen mechanism \cite{Friedrich1985}. 

Although the photonic structure of the M\"oR differs fundamentally from those studied previously \cite{Solodovchenko2025} in that it contains both topological and trivial photonic galleries, the mechanism for the formation of cascades of q-BIC, shown in Fig.~\ref{fig:fig02}, remains the same. Despite multiple anti-crossings in the regions where q-BICs exist, the photonic branches do not change their indices $(r,z,m)$, Fig.~\ref{fig:fig02}b;  therefore, we term such cascades ``ordinary''. This fact is significant for the discussion of ``topological'' q-BICs in the next section, where we will demonstrate the exchange of azimuthal indices $(m_{\mathrm{radial}} \leftrightarrow m_{\mathrm{axial}})$ following the anti-crossing of the branches.


\subsection*{Doublets of ``Topological'' q-BICs in the square cross-section region of M\"oRs.}\label{subsec2.4}

%
\begin{figure}[!htbp]
   \centering
   \includegraphics[width=\linewidth]{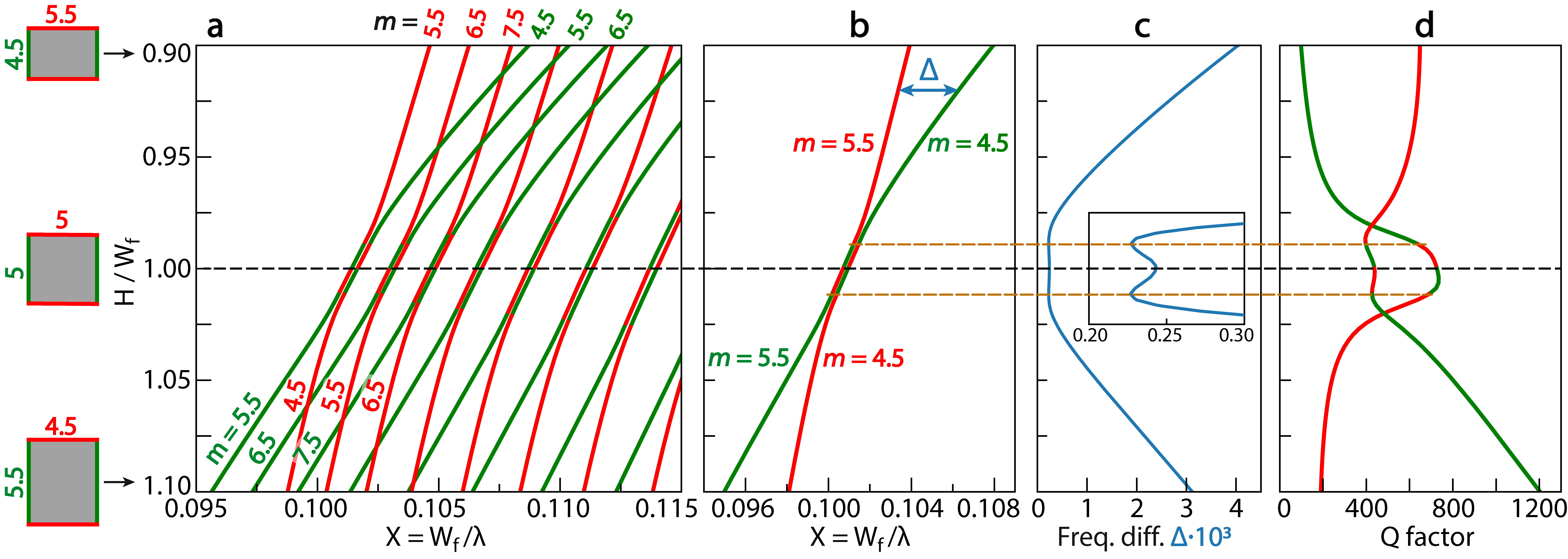}
   \caption
{\textbf{Gallery of paired topological q-BICs of M\"oRs in the region of a square cross-section.} \textbf{a}, Cascade of paired q-BICs formed by the radial (red) photonic gallery $(r = 1, z = 0, m)$ and axial (green) photonic gallery $(r = 0, z = 1, m)$ in the region of square cross-section of the M\"oR. On the left, three images illustrate the change in the cross-sectional shape of the M\"oR and the variation of the azimuthal indices $m$ for a pair of interfering modes $(m=4.5, \, 5.5)$ as its height $H$ is varied at a constant width $W_f$. \textbf{b}, Dependences of the frequencies of the modes $m=4.5, \, 5.5$ on an increased frequency scale.  \textbf{c}, The enlarged difference in frequencies between the two branches $m=4.5, \, 5.5$ demonstrates double anti-crossing (two minima between the branches) caused by two q-BICs at $\mathrm{H}/\mathrm{W}_\mathrm{f} \sim 0.99$ and $\mathrm{H}/\mathrm{W}_\mathrm{f} \sim 1.01$. \textbf{d}, Calculated $Q$ factors of the two modes in panel {b}. The two brown dashed lines in ({b-d}) correspond to red $\leftrightarrow$ green color exchanges caused by two q-BICs, by analogy with Fig.~\ref{fig:fig02}, while a third color exchange arises from a change in the ratio of the narrow and wide MöSs at $\mathrm{H}/\mathrm{W}_\mathrm{f} =1$. $\epsilon_{res}=40$, $\tan \delta_{res} =10^{-4}$, $\epsilon_{env}=1$.
}
\label{fig:fig03}
\end{figure}
%

Figure~\ref{fig:fig03} shows a new type of q-BICs, which we call paired ``topological'', since they are fundamentally different from the ``ordinary'' q-BICs in Fig.~\ref{fig:fig02}. The existence of q-BIC pairs follows from Fig.~\ref{fig:fig03}b-d. The Fig.~\ref{fig:fig03}c clearly reveals two minimum distances between the branches, corresponding to two anti-crossing regions (at $\mathrm{H}/\mathrm{W}_\mathrm{f} \sim 0.99$ and $\mathrm{H}/\mathrm{W}_\mathrm{f} \sim  1.01$). Further, in accordance with Friedrich-Wintgen's theory \cite{Friedrich1985}, one of the interfering branches should exhibit a minimum $Q$ factor, while the other demonstrates the primary q-BIC effect: a significant increase in the $Q$ factor. Indeed, Fig.~\ref{fig:fig03}d shows two distinct $Q$ minima and two $Q$ maxima, although the latter are less pronounced due to the overlap of two intense resonance profiles.

However, the most unexpected effect is the exchange of azimuthal indices $m$ in each pair of interfering modes at the square-cross-section point $\mathrm{H}/\mathrm{W}_\mathrm{f} = 1$. This effect consists of a change in the azimuthal indices $m$ along the radial (red, $5.5 \to 4.5$) and axial (green, $4.5 \to 5.5$) dispersion branches shown in Fig.~\ref{fig:fig03}b; it is also reflected in Fig.~\ref{fig:fig03}d by a red $\leftrightarrow$ green color exchange at $\mathrm{H}/\mathrm{W}_\mathrm{f} = 1$. This behavior differs fundamentally from that shown in Fig.~\ref{fig:fig02}a and previously observed for different structures \cite{Rybin2017,Bogdanov2019,Solodovchenko2021,Solodovchenko2025,Bochek2022}. As demonstrated below, the effect is associated with a unique exchange of fields on the M\"oSs during the transition through the square cross-section; therefore, we term such q-BICs ``topological''.


\subsection*{Two M\"oSs exchange fields in M\"oR}\label{subsec2.5}

%
\begin{figure}[!htbp]
   \centering
   \includegraphics[width=\linewidth]{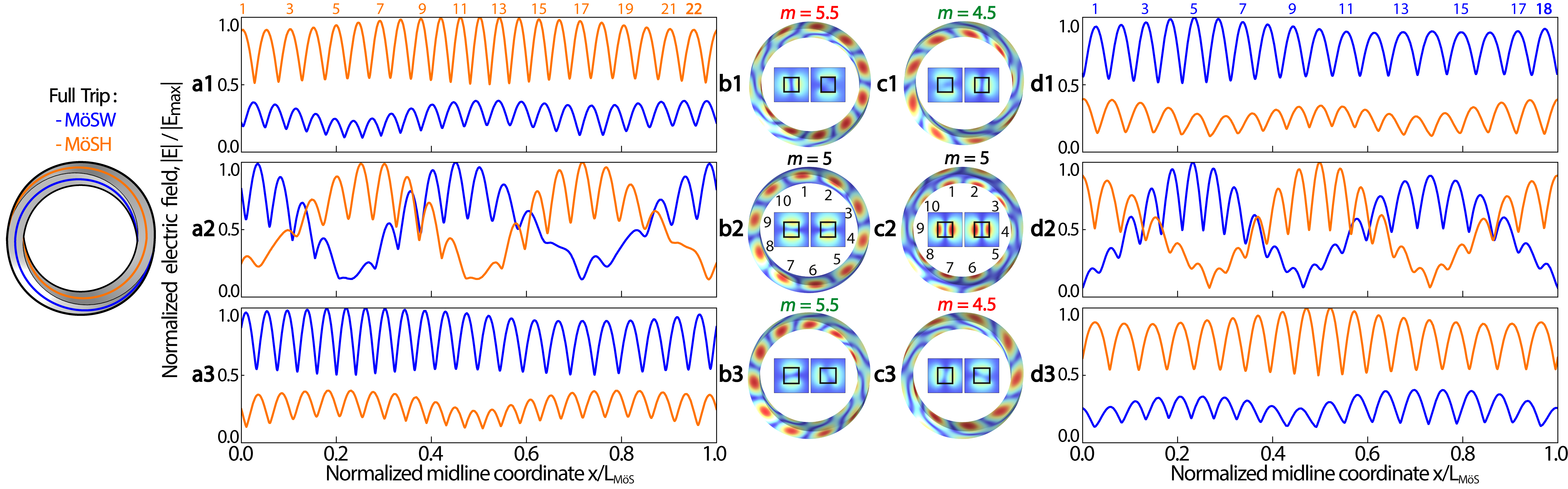}
   \caption
{\textbf{Exchange of fields between two M\"oSs in the area of a square cross-section of M\"oR.} Modes $m=5.5$ and $m=4.5$ are considered as an example. The colors correspond to the blue trajectory along the strip of variable width M\"oSW and the orange trajectory along the strip of constant width M\"oSH when walking along the midline of each of the two M\"oSs. A schematic representation is shown in the figure on the left. \textbf{a1-a3}, Distribution of the field $|\mathbf{E}|$ of the mode $m=5.5$ over two M\"oSs at $\mathrm{H}/\mathrm{W}_\mathrm{f} = 0.95, \, 1.00, \, 1.05$. \textbf{b1-b3}, \textbf{c1-c3}, Distribution of the field $\mathbf{E}$ on M\"oRs top view, depending on $\mathrm{H}/\mathrm{W}_\mathrm{f}$ for modes $m=5.5, \, 4.5$. In ({b2}) and ({c2}) $10$ clearly distinguishable maxima of the electric field are marked, corresponding to the mode index $m=5$. The inserts inside the resonators show the distributions of the $\mathbf{E}$ field in the resonator itself in a vertical section, as shown in enlarged scale in the Supplementary, Fig.~3s. In ({b2}) and the insert inside the resonator, the field is oriented along the upper and lower walls, while in ({c2}), the field is oriented along the side walls. Such a field distribution resembles the doubly degenerate phononic modes $\mathrm{E}_\mathrm{g}$ and $\mathrm{E}_\mathrm{u}$ in crystals of tetragonal symmetry, Fig.~3s. \textbf{d1-d3}, Field distribution for mode $m=4.5$ along two M\"oSs at $\mathrm{H}/\mathrm{W}_\mathrm{f} = 0.95, \, 1.00, \, 1.05$. For the mode $m=5.5$, $22$ maxima of the field amplitude $|\mathbf{E}|$ are numbered above the panel {a1}, for the mode $m=4.5$, $18$ maxima $|\mathbf{E}|$ are numbered above the panel {d1}. $\epsilon_{res}=40$, $\tan \delta_{res} =10^{-4}$, $\epsilon_{env}=1$.
}
\label{fig:fig04}
\end{figure}
%

Figure~\ref{fig:fig04} shows a spectacular mechanism of photonic branch transformation in the region of a square cross-section of M\"oR. Calculation of the normalized amplitude of the electric field $|\mathbf{E}|/|\mathbf{E}_\mathrm{max}|$ on the resonator surface during complete traverse along the middle of two M\"oSs allows us to uniquely determine the azimuthal index $m$. To determine the index $m$, one must account for the double traversal of the M\"oR over the entire surface and the doubling resulting from the field modulus, i.e. the number of $|\mathbf{E}|$ maxima must be divided by four. Therefore we get indices $m$ from Figs.~\ref{fig:fig04}a1,a3 as $22:4=5.5$ and from Figs.~\ref{fig:fig04}d1,d3 as $18:4=4.5$. Determining the index $m$ from the field image on the resonator itself is difficult due to the twisted cross-section. However, for a square cross-section, the field $\mathbf{E}$ along the M\"oR is oriented in an unexpected way and for both modes $10$ $\mathbf{E}$ field maxima are uniquely identified, which corresponds to an integer index $m = 5$ (Fig.~\ref{fig:fig04}b2,c2). Here we observe a unique picture: with the square cross-section and the equivalence of the two M\"oSs, the field is divided exactly in half between them, with the amplitudes alternating, changing twice along the M\"oS length (Fig.~\ref{fig:fig04}a2,d2). A more detailed transformation of fields when $\mathrm{H}/\mathrm{W}_\mathrm{f}$ changes is presented in the Supplementary, Fig.~4s.  

Now it is possible to explain the observation of a topological pair of q-BICs, Figs.~\ref{fig:fig03},~\ref{fig:fig04}. The effects are associated with a non-trivial transformation of the M\"oR at $\mathrm{H}/\mathrm{W}_\mathrm{f} =1$: the M\"oSH which was narrow in the thin M\"oR at $\mathrm{H}/\mathrm{W}_\mathrm{f} <1$, becomes wide at $\mathrm{H}/\mathrm{W}_\mathrm{f} >1$, and M\"oSW accordingly becomes narrow. As a result, in Fig.~\ref{fig:fig04} we observe a peculiar exchange of fields: in both photonic branches, the intense field passes from one stripe to another, the distribution of fields in the $5.5$ and $4.5$ modes swaps ($\mathrm{a1} \to \mathrm{d3}$ and $\mathrm{d1} \to \mathrm{a3}$, Fig.~\ref{fig:fig04}). Thus, for both $\mathrm{H}/\mathrm{W}_\mathrm{f} < 1$ and $\mathrm{H}/\mathrm{W}_\mathrm{f} > 1$, a larger number of resonant mode waves $(m)$ is accommodated on the M\"oS with the larger area. As a result, the well-known rule for "ordinary" q-BICs (Fig.~\ref{fig:fig02} and \cite{Rybin2017,Bogdanov2019,Solodovchenko2021,Solodovchenko2025,Bochek2022}): the preservation of the dispersion dependences of the two modes after passing through the anti-crossing region is violated. We demonstrate a fundamentally different picture regarding the exchange of longitudinal indices $m$ $(4.5\leftrightarrow5.5)$ between two interfering branches in the anti-crossing region (Figs.~\ref{fig:fig03},~\ref{fig:fig04}), driven by the nontrivial topological structure of the M\"oR.

\FloatBarrier
\section*{Discussion}\label{sec3}

%
\begin{figure}[!htbp]
   \centering
   \includegraphics[width=0.5\linewidth]{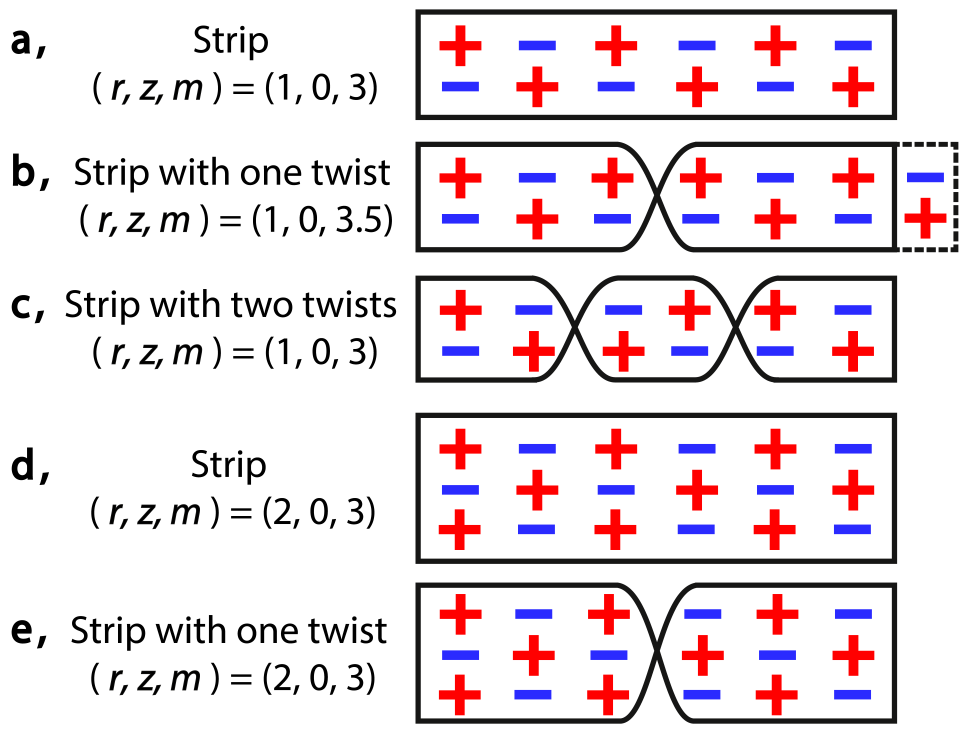}
   \caption
{\textbf{Schematic: dependence of BPh on the sum of mode indices $(r+z)$ and the number of twists.} The figure shows the distribution of the electromagnetic field in thin planar dielectric strips, as well as in strips with one and two twists for the eigenmodes $(r=1,z=0,m=3)$ ({a},{c}) and $(r=2,z=0,m=3)$ ({d},{e}). \textbf{b}, Thin dielectric strip with one twist and distribution of electromagnetic field for the eigenmode $(1,0,3)$. On the right, a dotted line adds half a period, which increases the mode index to $(1,0, 3.5)$. The plus and minus signs reflect the amplitude of the electromagnetic wave taking into account the phase.
}
\label{fig:fig05}
\end{figure}
%

Our conclusion regarding the spectral dependence of BPh has a vivid graphical interpretation, Fig.~\ref{fig:fig05}. We consider what happens when the ends of a thin dielectric strip with a rectangular cross-sections are joined, that is, when creating a planar RR ({a},{d}), a M\"oR with one twist ({b},{e}), and a M\"oR with two twists ({c}). Connecting two ends of a stripe with opposite phases leads to the creation of a RR ({a},{d}) and a M\"oR ({c},{e}) with an azimuthal index $m=3$. Connecting two ends of a stripe with the same phases ({b}) results in a twofold increase in a single period only and disrupts the overall periodicity in a closed stripe. Therefore, to obtain a resonant mode in the M\"oR with one twist, we must add, as shown by the dotted line in ({b}), or ``cut off'' half the period and obtain, respectively, a M\"oR with a half-integer azimuthal index of $3.5$ or $2.5$.  

It is clear that such reasoning is valid for any values of the $(r,z,m)$ modes and any number of twists. Thus, this simple scheme fully confirms our main conclusion about the spectral dependence of the BPh, which in a thin M\"oR with an odd number of twists is $\pi$ for an odd value of $(r+z)$ and is zero for an even value of $(r+z)$.

%
\begin{figure}[!htbp]
   \centering
   \includegraphics[width=\linewidth]{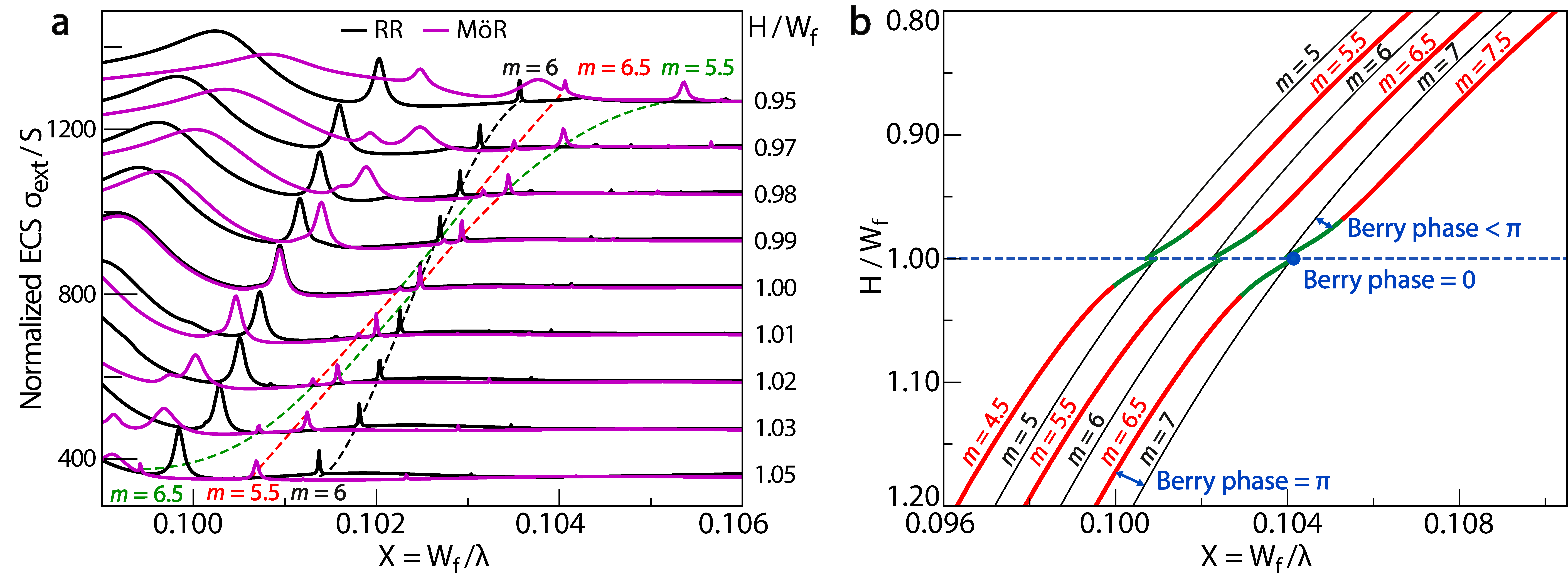}
   \caption
{\textbf{Resonant topology: BPh as a function of the difference between the eigenmodes of M\"oR and planar RR.} The dependences of the longitudinal modes of the first radial gallery of a planar RR with $m=5, \, 6, \, 7$ are shown in black. The dependences of the longitudinal modes of the first radial gallery of a M\"oR with half-integer indices $m$ indicated in the figure are shown in red and green. Both resonators have the same cross-section, midline length, and permittivity, as in Fig.~\ref{fig:fig03}. The change in colors red $\leftrightarrow$ green for the MöRs branches is associated with q-BICs (Fig.~\ref{fig:fig03}b). The slight mismatch of the M\"oRs frequencies at $\mathrm{H}/\mathrm{W}_\mathrm{f}=1$ (red and green curves) is due to the anti-crossing effect in the q-BIC region. The slight mismatch of the RRs frequencies at $\mathrm{H}/\mathrm{W}_\mathrm{f}=1$ (two black curves) is due to the different curvatures of the two lateral and two parallel planes of the RR (Supplementary, Fig.~5s). $\epsilon_{res}=40$, $\tan \delta_{res} =10^{-4}$, $\epsilon_{env}=1$.
}
\label{fig:fig06}
\end{figure}
%

The set of dependences of the RR and M\"oR eigenmodes on the parameter $\mathrm{H}/\mathrm{W}_\mathrm{f}$ defines an elegant spectroscopic method for observing the transformation of the BPh, including its disappearance in a square M\"oR (Fig.~\ref{fig:fig06}). In thin MöR, the frequency of the topological mode $m+1/2$ lies midway between the frequencies of the trivial modes $m$ and $m+1$, which corresponds to BPh of $\pi$ (Fig.~\ref{fig:fig01}). This relative positioning of the modes is maintained up to the immediate vicinity of the M\"oR's square cross-section (Fig.~\ref{fig:fig06}b), where the region of softening of MöR’s modes is observed. Assuming that the BPh magnitude is proportional to the frequency difference between the RR and M\"oR modes (Fig.~\ref{fig:fig06}), we see that as $\mathrm{H}/\mathrm{W}_\mathrm{f}\rightarrow1$, this frequency difference tends toward zero, which corresponds to BPh $=0$. From Fig.~\ref{fig:fig06}, the region of «softening» of the BPh can be estimated as $\sim 0.05\mathrm{H}/\mathrm{W}_\mathrm{f}$.

In 1956, S. Pancharatnam, within the framework of classical optics \cite{Pancharatnam1956}, and later M.V. Berry, from the standpoint of quantum mechanics \cite{Berry1984}, described an additional contribution to the well-known dynamical phase of a propagating wave, which M.V.Berry called ``a geometrical phase factor''. In this work, we extend the concept of the geometrical BPh by experimentally and theoretically demonstrating its dependence on the symmetry of photonic modes. In the extinction spectra of a single-twist M\"oR, we observe two types of modes: first, the expected topological modes $(1,0,m)$ with a BPh of $\pi$ (Fig.~\ref{fig:fig01}); and secondly, for the first time in thin M\"oR, we demonstrate modes with zero BPh: modes with $(2,0,m)$ symmetry (Fig.~\ref{fig:fig01}) and $(0,2,m)$ symmetry (Fig.~\ref{fig:fig02}). 

The discovered effects fundamentally expand the possibilities for manipulating the BPh, which determines a wide range of physical phenomena, including various Hall effects, ferroelectricity, quantum charge pumping, and orbital magnetism \cite{Niu1985,Resta1998,Xiao2010,Tuloup2023,Jin2024,Barge2026}.

\backmatter


\bmhead{Acknowledgments}

The authors acknowledge the financial support from the Russian Science Foundation No 23-12-00114-\textcyr{П}, https://rscf.ru/project/23-12-00114-\textcyr{П}/.

The authors thank A. Bogdanov, K. Friziuk and K. Koshelev for discussions of the work and helpful comments.




\FloatBarrier
\section*{Methods}\label{sec4}

\FloatBarrier
\subsection*{Sample fabrication}\label{subsec4.2}

The planar RR and the M\"oR were fabricated by 3D printing at ITMO University using a
FlyingBear Reborn2 printer and Zetamix Epsilon 10 filament (Ref.\cite{Filament}).
Both resonators had a \(20\,\mathrm{mm}\times4\,\mathrm{mm}\) rectangular
cross-section and the middle line radius of \(40\,\mathrm{mm}\). The dimensions were
chosen to place the longitudinal resonances of interest in the microwave
range.

The nozzle and build plate were held at approximately \(280\,^\circ\mathrm{C}\)
and \(110\,^\circ\mathrm{C}\), respectively. The printing speed was
\(2.5\,\mathrm{mm\,s^{-1}}\) for the first layer and
\(15\,\mathrm{mm\,s^{-1}}\) for subsequent layers. Comparing measured and
calculated extinction spectra gave an effective relative permittivity of
\(\epsilon_{res}=8.23\) and a loss tangent of
\(\tan\delta_{res}=1.1\times10^{-4}\) for the printed material. Additional
fabrication details are given in Sec.~1 of the Supplemental Material.

\FloatBarrier
\subsection*{Numerical simulation}\label{subsec4.1}

We used COMSOL Multiphysics to calculate extinction spectra and to track the
resonant frequencies, electric-field profiles, and $Q$ factors as the
resonator geometry was varied. The extinction cross-section spectra were obtained from
frequency-domain scattering simulations. The resonator was placed in a
spherical vacuum domain surrounded by a perfectly matched layer (PML). A
scattering boundary condition was applied at the outer PML-boundary to reduce
spurious reflections.

For the spectra in Fig.~\ref{fig:fig01}, the incident wave propagated along the \(x\) axis
and was polarized along the \(y\) axis. For the calculated spectra in
Fig.~\ref{fig:fig06}a, the wave vector was tilted in the \(xz\) plane while the electric
field remained parallel to the \(y\) axis. This configuration made additional
resonances visible in the extinction spectra. The simulation domain and both
excitation geometries are shown in Fig.~1s of the Supplemental Material, Sec. 2.

\FloatBarrier
\subsection*{Methodology for measuring extinction spectra}\label{subsec4.3}

Microwave extinction cross-section spectra were measured in an anechoic chamber at ITMO
University using a two-port vector network analyzer (Planar S5085) connected
to a pair of horn antennas (Fig.~6s).
The sample was placed between the transmitting and receiving antennas.
The complex forward transmission coefficient was recorded, first without the sample (\(S_{21}^{b}\))
and then with the sample in place (\(S_{21}^{s}\)).

The extinction cross section was obtained from the forward transmission
using the optical theorem:
\[
\sigma_{\mathrm{ext}}(\omega)
  = \frac{4\pi}{k}\frac{L_1L_2}{L_1+L_2}
    \operatorname{Im}\!\left(\frac{S_{21}^{s}}{S_{21}^{b}}\right).
\]
Here \(k=\omega/c\) is the incident wavenumber in air, and \(L_1\) and \(L_2\)
are the distances from the sample to the transmitting and receiving antennas,
respectively. For comparison, the measured and calculated extinction cross
sections were divided by the same reference area, \(S=2R_{\mathrm{out}}h\),
the projected area of the planar RR viewed edge-on, where \(R_{\mathrm{out}}\)
is its outer radius and \(h\) its height. Additional
measurments details are given in Sec.~3 of the Supplemental Material.

\FloatBarrier
\subsection*{Theoretical calculation of the Berry phase taking into account the symmetry of photon modes}\label{subsec4.4}

We model the M\"oR and the reference RR as dielectric waveguides with the same circular centerline of radius $R$, rectangular cross-section $2a\times2b$, and resonator and environment permittivities $\epsilon_{res}=n^2$ and $\epsilon_{res}>\epsilon_{env}$ respectively. Both media are isotropic and nonmagnetic. We neglect material absorption and dispersion and assume that only a small fraction of the guided power is lost to radiation over one circuit. We use coordinates $(s,u,v)$ comoving with the cross-section, with $s$ the arclength along the centerline and the sidewalls at $u=\pm a$ and $v=\pm b$. The material axes rotate relative to the Frenet--Serret frame through an angle $\Omega s$, where $\Omega=N\pi/L$, $L=2\pi R$, and $N$ is the number of twists~\cite{Yabe1984,Takagi1992,Fomin2025}. The endpoint identification is $(s+L,u,v)\sim(s,Pu,Pv)$, with $P=(-1)^N$. We denote the transverse coordinate map by $\tau_N(u,v)=(Pu,Pv)$ and its action on the field by $\mathcal T_N$.

We write the source-free Maxwell equations in the material frame and impose the dielectric interface conditions on the sidewalls. Eliminating the longitudinal field components gives a propagation equation for $F=(E_u,E_v,H_u,H_v)^T$ with operator $B_0+\Omega J+K(s)$, as derived in Sec.~4 of the Supplementary. Here $B_0$ is the straight-guide operator, $J$ generates rotations of the transverse profiles and their vector components, and $K(s)$ is the curvature correction. We project this equation onto the bound modes of the straight guide, normalized to unit longitudinal power. The single-mode approximation requires the transverse extent of the field to be small compared with $R$ and the selected mode to remain separated from cutoff. For each coupled mode, the twist and curvature matrix elements must be small compared with the difference in propagation constants. Reflection symmetry makes both diagonal matrix elements vanish. At fixed $N$ and cross-section, the resulting correction to the dynamical phase over one circuit is $O(R^{-1})$ for an isolated mode. We use this approximation away from the polarization degeneracy near the square cross-section.

We assign the gallery indices $r,z$ from the sign changes of a transverse electric component along the width and height. With the nodal convention given in Sec.~4 of the Supplementary, its spatial parity under $(u,v)\mapsto(-u,-v)$ is $(-1)^{r+z-1}$. A half-turn also reverses the transverse vector component, giving the mode the half-turn eigenvalue $p_{rz}=(-1)^{r+z}$. We determine the allowed longitudinal indices by imposing single-valuedness after one circuit, $e^{iq_{mrz}L}p_{rz}^N=1$, which gives
\begin{equation}
    q_{mrz}=\frac{m}{R},\qquad
    m\in\mathbb Z+\frac{N(r+z)}{2}.
    \label{eq:eq01}
\end{equation}

To calculate the BPh, we express the normalized transverse profile $\mathbf w_{rz}(s)$ in the periodic Frenet frame and factor out the longitudinal factor $e^{iq_{mrz}s}$. In terms of the power-flux form $\langle\cdot,\cdot\rangle_P$ integrated over the transverse directions, the phase is
\begin{equation}
    \gamma_{rz}
    =i\int_0^L
    \langle\mathbf w_{rz}(s),\partial_s\mathbf w_{rz}(s)\rangle_P\,ds
    +\arg\langle\mathbf w_{rz}(0),\mathbf w_{rz}(L)\rangle_P
    \pmod{2\pi}.
    \label{eq:eq02}
\end{equation}
For a profile transported with the cross-section, $\partial_s\mathbf w_{rz}=-i\Omega J\mathbf w_{rz}$. The diagonal element of $J$ vanishes by reflection symmetry, so only the endpoint overlap contributes. Using $\mathbf w_{rz}(L)=\mathcal T_N\mathbf w_{rz}(0)=p_{rz}^N\mathbf w_{rz}(0)$ gives
\begin{equation}
    \gamma_{rz}=N\pi(r+z)\pmod{2\pi},\qquad
    q_{mrz}L+\gamma_{rz}=2\pi\ell,\qquad \ell\in\mathbb Z.
    \label{eq:eq03}
\end{equation}

For the spectral comparison in Fig.~\ref{fig:fig01}, we use adjacent RR resonances of the same gallery. Their propagation phases differ by $2\pi$ and their angular frequencies by the local free spectral range $\Xi$. For odd $N(r+z)$, Eq.~\eqref{eq:eq03} places the M\"obius resonance at a phase increase of $\pi$ from the lower RR resonance. We convert this to a frequency shift of $\Xi/2$ when the straight-guide dispersion $q(\omega)$ is approximately linear over that interval.

\end{document}


\title[Supplemental Material]{Spectral dependence of Berry phase \\ and topological bound states in the continuum in M\"obius resonators}
\subtitle{\textbf{Supplemental Material}}

\author[]{\fnm{Mikhail E.} \sur{Bochkarev}}\email{mikhail.bochkarev@metalab.ifmo.ru}
\equalcont{These authors contributed equally.}

\author[]{\fnm{Timur Z.} \sur{Seidov}}\email{tim.phys.math@gmail.com}
\equalcont{These authors contributed equally.}

\author[]{\fnm{Nikolay S.} \sur{Solodovchenko}}\email{n.solodovechenko@metalab.ifmo.ru}
\equalcont{These authors contributed equally.}

\author[]{\fnm{Kirill B.} \sur{Samusev}}\email{k.samusev@metalab.ifmo.ru}

\author*[]{\fnm{Mikhail F.} \sur{Limonov}}\email{m.limonov@metalab.ifmo.ru}

\affil[]{\orgdiv{School of Physics and Engineering}, \orgname{ITMO University}, \\ \orgaddress{\street{Kronverksky Pr. 49}, \city{St. Petersburg}, \postcode{197101}, \country{Russia}}}


\maketitle

\FloatBarrier
\section{Creation of dielectric samples of ring resonators and Möbius resonators}

To fabricate samples of the ring resonator (RR) and Möbius resonator (MöR) with one twist, we used ZetaMix epsilon 10 filament, an ABS matrix containing 80\% titanium dioxide (Ref. {[}39{]} in the main text: \url{https://zetamix.fr/produit/filament-epsilon/}). Since 3D printing involves creating layers using a plastic filament that follows a ``labyrinthine-like'' path, it was possible to achieve the effective dielectric permittivity of $\epsilon_{res}=8.23$ and the dielectric loss tangent of $\tan\delta_{res}=1.1\times10^{-4}$ in $4.5-10.5$ GHz range. This permittivity was obtained by comparison of experimental and calculated spectra.

The sample printing was performed at ITMO University (St. Petersburg) on a FlyingBear Reborn2 3D-printer, which operates on the principle of layer-by-layer creation of a three-dimensional object according to a given program. Due to the fragility of the material due to the high titanium dioxide content, the spool with the plastic was suspended at a height of one meter above the 3D printer. Printing mode: nozzle temperature approximately $280\,{}^\circ\mathrm{C}$, heating platform temperature $110\,{}^\circ\mathrm{C}$. According to the recommendations of the Plastic Filament Manufacturer {[}39{]}, the material application speed was 2.5 mm/s for the first layer and 15 mm/s for subsequent layers. As a result, two types of resonators were manufactured: Ring resonators and Möbius resonators with one twist (Fig.~1 in the main text) which required 6 and 13 hours to print, respectively. The increase in printing time is explained by the need to print additional support structures for more complex resonator geometry. Both resonators have a constant cross-section. The resonator geometry was modeled by rotating a rectangular cross-section with a width $W=20\,\mathrm{mm}$ and a height $h=4\,\mathrm{mm}$ along a centerline with a radius $R_{\mathrm{central}}$ = 40 mm, which corresponds to an inner-to-outer radius ratio of $R_{\mathrm{in}}$/$R_{\mathrm{out}}$ = 0.6. The dimensions were chosen such that the longitudinal modes of the resonators fell in the frequency range from 4.5 to 10.5 GHz.

\FloatBarrier
\section{Numerical simulation}

The calculations were performed using the COMSOL Multiphysics package. With eigensolver module we obtained the evolution of eigenvalues (resonant wavelength or frequency), eigenstates (field distributions) and quality factor during shape transformation. 

\begin{figure}[H]
  \centering
  \includegraphics[width=.50\linewidth]{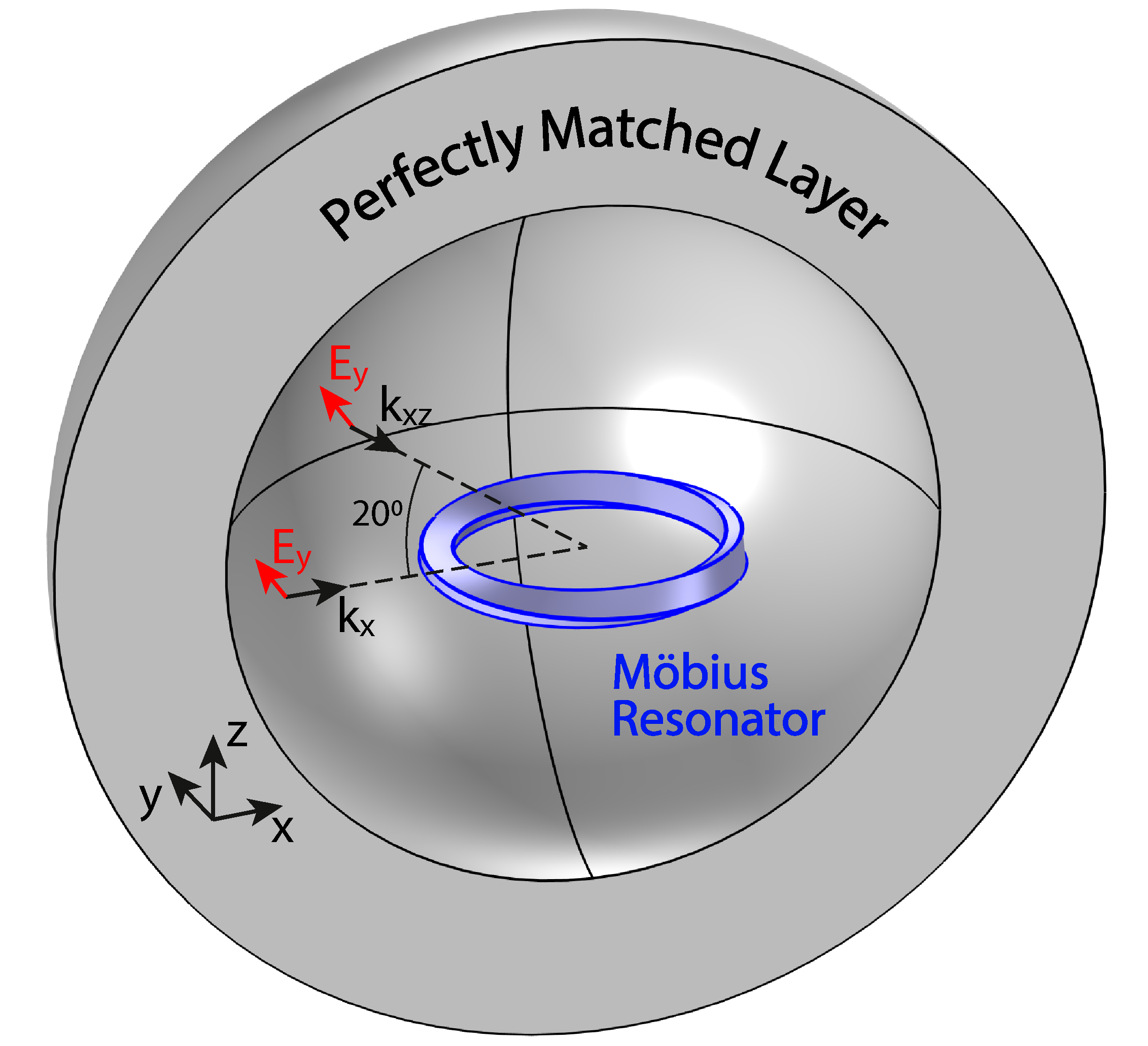}
  \caption{\textbf{Schematic representation of the model.} The gray outer shell is the PML layer. The first excitation geometry with ($k_x$, $E_y$) was used in the experiment and calculations shown in Fig.~1 of the main text, the second excitation geometry with ($k_{xz}$, $E_y$) was used in calculations for Fig.~6 to observe bigger number of resonances in spectra.}
  \label{fig:1s}
\end{figure}

The frequency domain solver module was used for calculation of the far-field ECS. All calculations were performed for the two excitation geometries, which are shown in (Fig.~\ref{fig:1s}). In the simulation volume the resonator was surrounded by a vacuum sphere which ends by a spherical PML layer with scattering boundary condition on the outer surface to escape parasitic reflections back to the simulation volume (Fig.~\ref{fig:1s}). Numerical and experimental extinction cross sections (ECS, Fig.~1 in the main text) were normalized to the area of the ring's geometric shadow, $S=2hR_{\mathrm{out}}$.

\begin{figure}[H]
  \centering
  \includegraphics[width=.80\linewidth]{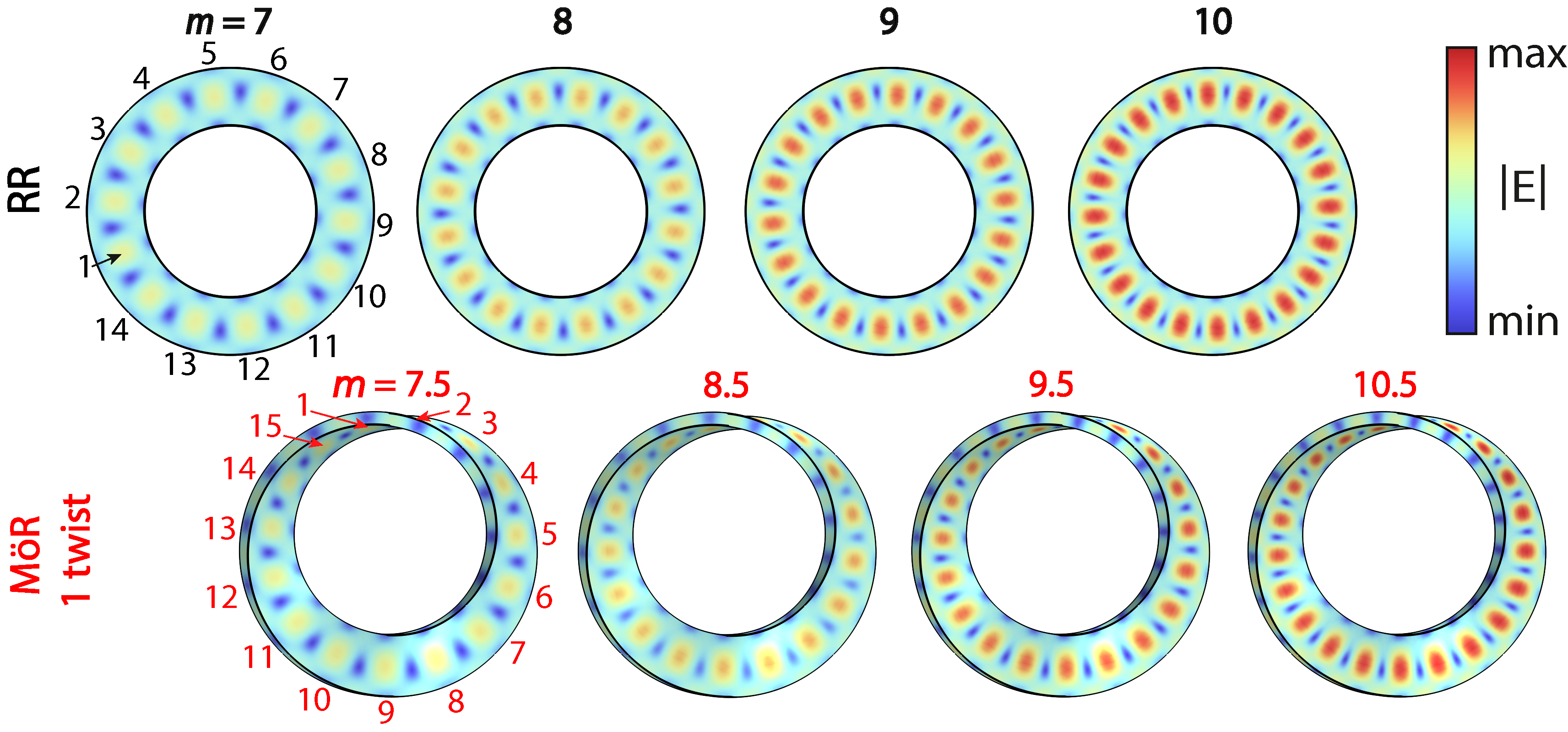}
  \caption{\textbf{Distribution of the field $\lvert\mathbf{E}\rvert$ in a RR and a MöR for the modes of the first gallery on an enlarged scale compared to Fig.~1 in the article.}}
  \label{fig:2s}
\end{figure}

\begin{figure}[H]
  \centering
  \includegraphics[width=.80\linewidth]{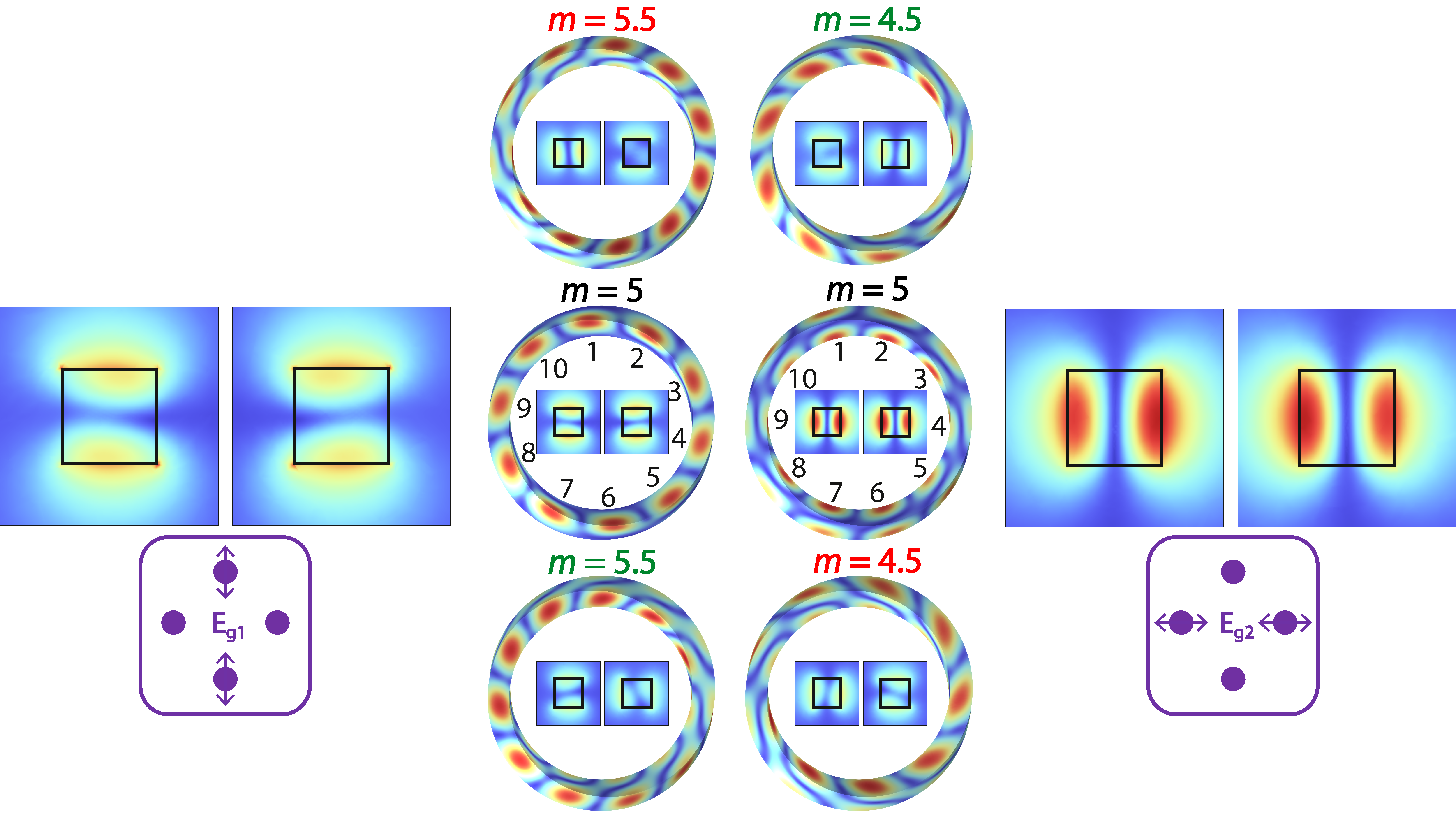}
  \caption{\textbf{Distribution of the field $\lvert\mathbf{E}\rvert$ in a MöR for modes $m=5.5$ and $m=4.5$ on an enlarged scale compared to Fig.~4 in the article.} We draw attention to the field distribution inside and around the resonators, shown in the insets inside the resonators. Of particular interest is the field distribution for resonators with a square cross-section ($m=5$). For the low-frequency mode, the field is distributed strictly on the outer surface of the resonator (viewed from above), while for high-frequency, it is distributed strictly along the sides of the resonator, allowing us to unambiguously determine the $m$ index in both cases. The purple frames show atomic displacements in crystals of tetragonal symmetry corresponding to doubly degenerate even phonon modes of $E_{\mathrm{g}}$ symmetry. A similar pattern applies to atomic displacements corresponding to doubly degenerate odd phonon modes of $E_{\mathrm{u}}$ symmetry.}
  \label{fig:3s}
\end{figure}

In three-dimensional planar RR and MöR, the eigenfunctions can be characterized by azimuthal ($m$) and two transverse mode indices along the width ($r$) and height ($z$) of the cross section, forming an ordered triple ($r,z,m$). Figures~\ref{fig:2s},~\ref{fig:3s} show examples of calculating the distribution of the $\mathbf{E}$ field in a ring resonator and a Möbius resonator on an enlarged scale.

\begin{figure}[H]
  \centering
  \includegraphics[width=.80\linewidth]{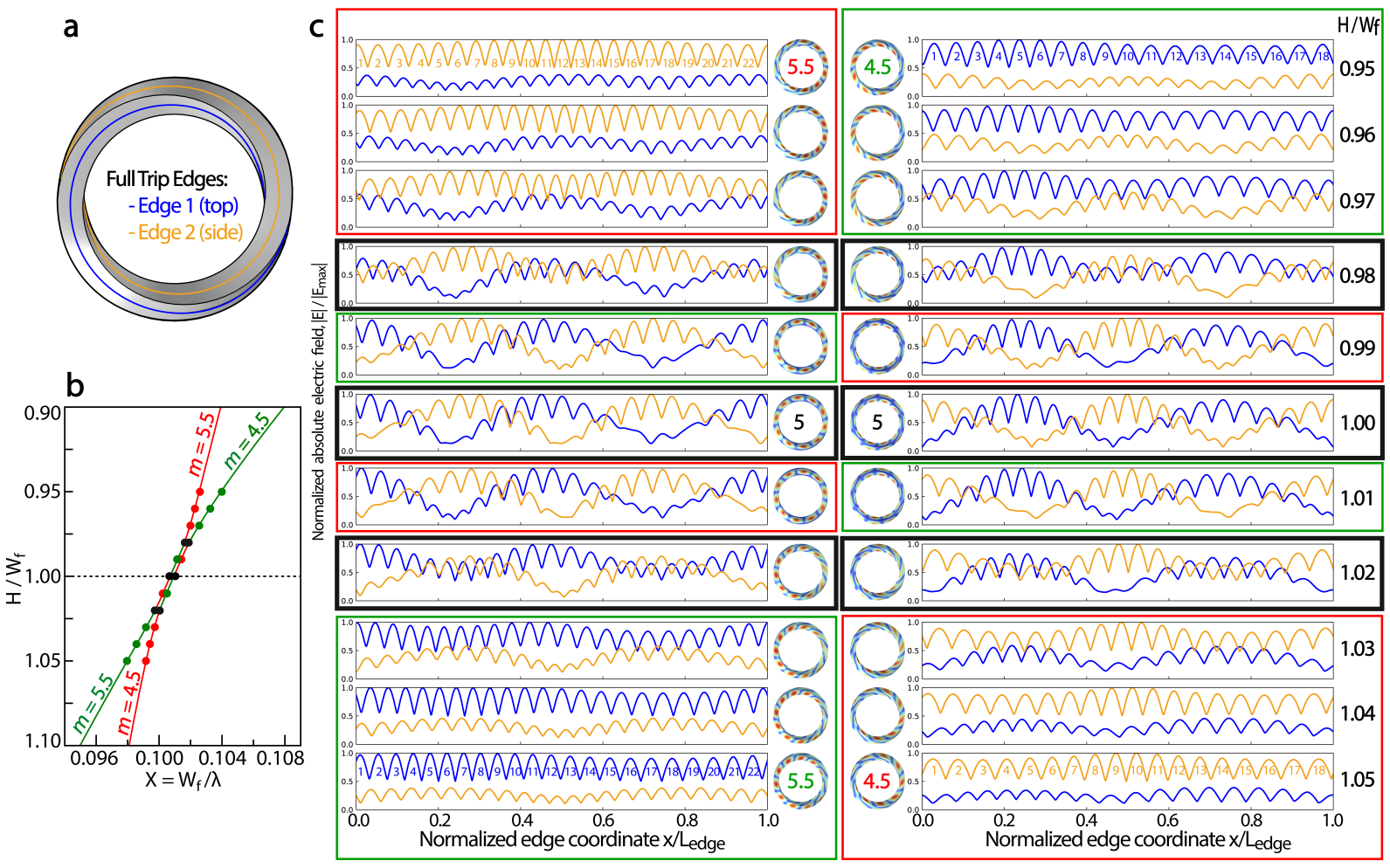}
  \caption{\textbf{Detailed changes in the field distribution across two surfaces of the MöS at 11 points as we move along the dispersion branches in the anticrossing and q-BIC region.} (a) The MöS with the orange and the blue middle lines for each of the two surfaces of the resonator. (b) Dispersion branches near the q-BIC region with $H/W_{\mathrm{f}}$ = 1. (c) Electric field distribution across the middle lines for the particular points marked in (b) for $H/W_{\mathrm{f}}$ from 0.95 to 1.05 with the step 0.01. Figure~4 in the article shows only three boxed field distributions for points $H/W_{\mathrm{f}}$ = 0.95, 1.00, and 1.05.}
  \label{fig:4s}
\end{figure}

\begin{figure}[H]
  \centering
  \includegraphics[width=.68\linewidth]{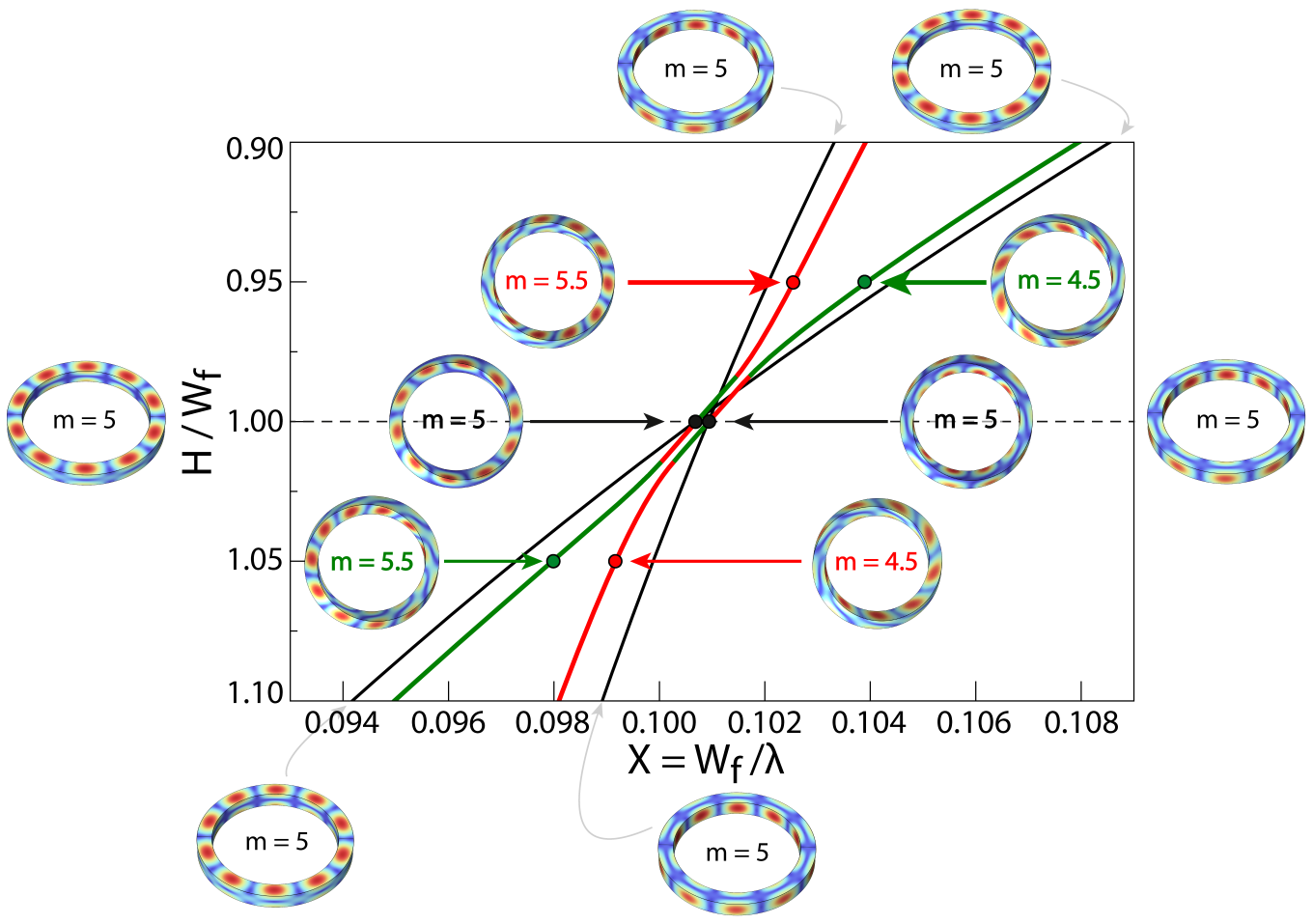}
  \caption{\textbf{Behavior of radial and axial modes RR (black lines) and MöR (green and red lines) in the region of a square cross-section.}}
  \label{fig:5s}
\end{figure}

We attribute the slight mismatch between the frequencies of the radial and axial modes of a flat RR with a square cross-section, $H/W_{\mathrm{f}}$ = 1 (the intersection of the two black curves with the horizontal line corresponding to $H/W_{\mathrm{f}}$ = 1 at different points, i.e., at slightly different frequencies) to the curvature of the RR, which affects the radial modes to a greater extent than the axial ones.

\FloatBarrier
\section{Methodology for measuring extinction spectra}

\begin{figure}[H]
  \centering
  \includegraphics[width=.72\linewidth]{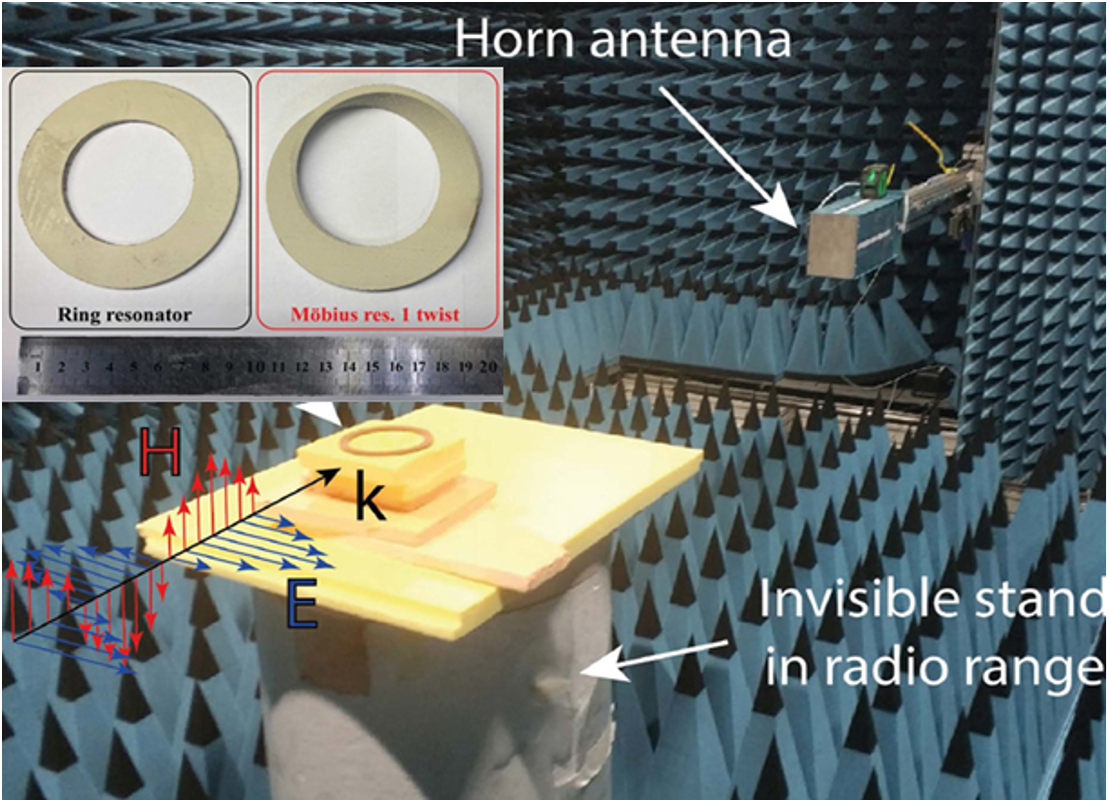}
  \caption{\textbf{Photograph of the anechoic chamber at ITMO University, where microwave experiments were conducted to measure the extinction spectra of a ring resonator and a Möbius resonator.}}
  \label{fig:6s}
\end{figure}

The microwave experimantal setup allowed us to measure the extinction cross section using the optical theorem in the anechoic chamber of ITMO University (Fig.~\ref{fig:6s}). The measurements were performed using a two-port vector network analyzer (Planar S5085), to which two horn antennas (TRIM 0.75--18 GHz) forming a quasi-plane wave were connected, installed at a distance of four meters from each other: $L_1$ = $L_2$ = 2 m -- the distance between source\textbackslash receiver antenna and the sample. The frequency range from 4.5 to 10.5 GHz was scanned using 16001 frequency points. The \(S_{21}\) parameter of the scattering matrix was measured twice: first in free space \(S_{21}^{b}\), and then with a sample \(S_{21}^{s}\) placed between the antennas. According to the optical theorem, the imaginary part of the forward scattering amplitude is proportional to the effective scattering cross section. Thus, the extinction cross section was calculated using the formula

\begin{equation*}
\sigma_{\mathrm{ext}}(\omega)
  = \frac{4\pi}{k}\operatorname{Im}\!\left(\frac{S_{21}^{s}}{S_{21}^{b}}\right)
    \left(\frac{L_1L_2}{L_1+L_2}\right),
\end{equation*}

where $k$ is the wave vector of the incident wave. The extinction cross sections obtained in both the calculations and the experiment were normalized to the geometric shadow of the RR, defined as $S=2hR_{\mathrm{out}}$.

\FloatBarrier
\section{Theoretical description of geometric phase and spectral shifts in a dielectric M\"obius resonator}
\label{sup:mobius}

\subsection{Geometry}
\begin{figure}[t]
\centering
\includegraphics[width=.68\linewidth]{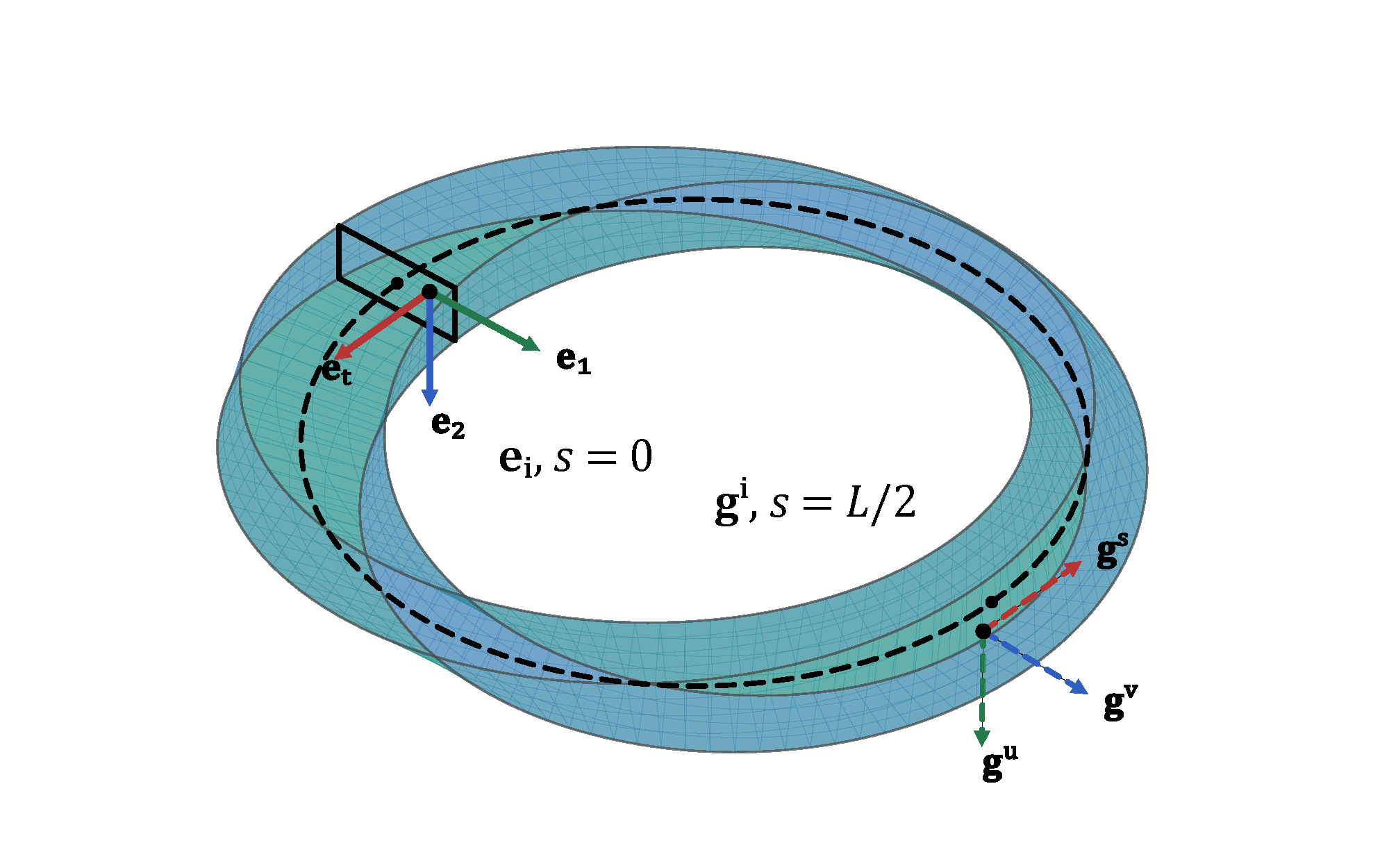}
\caption{Geometry of the M\"obius resonator for $N=1$. The material basis at $s=0$ and the reciprocal basis at $s=L/2$ are shown.}
\label{sup:mobius:fig:geometry}
\end{figure}

Let $\boldsymbol\gamma(s)$, $0\leq s\leq L$, be the circular centerline parametrized by arclength. Its Frenet--Serret frame $\{\mathbf t,\mathbf n,\mathbf b\}$ satisfies
\begin{equation}
\mathbf t'=\kappa\mathbf n,\qquad
\mathbf n'=-\kappa\mathbf t,\qquad \mathbf b'=0,
\end{equation}
where a prime denotes differentiation with respect to $s$ and the curvature is $\kappa=2\pi/L$ (Fig.~\ref{sup:mobius:fig:geometry}). We introduce a material frame $\mathbf e_i$ that rotates uniformly through $N\pi$ over one round trip, where the integer $N$ counts half-turns,
\begin{equation}\label{sup:mobius:eq:material-frame}
\mathbf e_t=\mathbf t,\quad
\mathbf e_1=\cos\vartheta\,\mathbf n+\sin\vartheta\,\mathbf b,\quad
\mathbf e_2=-\sin\vartheta\,\mathbf n+\cos\vartheta\,\mathbf b,\quad
\vartheta(s)=\Omega s,\quad \Omega=\frac{N\pi}{L}.
\end{equation}
At $s=L$, $\mathbf e_1(L)=P\mathbf e_1(0)$ and $\mathbf e_2(L)=P\mathbf e_2(0)$, where $P=(-1)^N$. The resonator bulk is parametrized by
\begin{equation}
\mathbf r(s,u,v)=\boldsymbol\gamma(s)+u\mathbf e_1(s)+v\mathbf e_2(s),
\qquad -a\leq u\leq a,\qquad -b\leq v\leq b.
\end{equation}
The full resonator width and height are $W=2a$ and $H=2b$, with aspect ratio $\eta=b/a=H/W$. The endpoints are identified by the transition map
\begin{equation}\label{sup:mobius:eq:seam}
(L,u,v)\sim(0,Pu,Pv),
\end{equation}
because $\mathbf r(L,u,v)=\mathbf r(0,Pu,Pv)$. For odd $N$, the mid-surface $v=0$ is a nonorientable M\"obius strip, whereas the three-dimensional dielectric domain is orientable because the transverse gluing map has determinant $P^2=1$. The modal closure phase is determined by the action of the twist on a particular vector mode.

Differentiating Eq.~\eqref{sup:mobius:eq:material-frame} yields
\begin{equation}
\mathbf e_1'=-\kappa\cos\vartheta\,\mathbf t+\Omega\mathbf e_2,
\qquad \mathbf e_2'=\kappa\sin\vartheta\,\mathbf t-\Omega\mathbf e_1.
\end{equation}
Consequently, the covariant coordinate basis vectors $\mathbf g_i=\partial_i\mathbf r$, with $i\in\{s,u,v\}$, are
\begin{align}
\mathbf g_s&=f\mathbf t-\Omega v\mathbf e_1+\Omega u\mathbf e_2,
&\mathbf g_u&=\mathbf e_1,&\mathbf g_v&=\mathbf e_2,\label{sup:mobius:eq:covariant-basis}\\
&f(s,u,v)=1-\kappa\bigl(u\cos\vartheta-v\sin\vartheta\bigr)&.
\end{align}
Here $R=L/(2\pi)$. The condition $R^2>a^2+b^2$ ensures $f>0$ throughout the resonator, and the same chart describes the exterior only within a regular tubular neighborhood. The metric and its determinant are
\begin{equation}
(g_{ij})=\begin{pmatrix}
f^2+\Omega^2(u^2+v^2)&-\Omega v&\Omega u\\
-\Omega v&1&0\\ \Omega u&0&1
\end{pmatrix},\qquad \sqrt g=f.
\end{equation}
The reciprocal basis $\mathbf g^i\cdot\mathbf g_j=\delta^i_j$ is
\begin{equation}
\mathbf g^s=\frac{\mathbf t}{f},\qquad
\mathbf g^u=\mathbf e_1+\frac{\Omega v}{f}\mathbf t,\qquad
\mathbf g^v=\mathbf e_2-\frac{\Omega u}{f}\mathbf t.
\end{equation}
The inverse metric is
\begin{equation}
(g^{ij})=\frac1{f^2}\begin{pmatrix}
1&\Omega v&-\Omega u\\
\Omega v&f^2+\Omega^2v^2&-\Omega^2uv\\
-\Omega u&-\Omega^2uv&f^2+\Omega^2u^2
\end{pmatrix}.
\end{equation}

\subsection{Maxwell equations and straight-guide modes}

The M\"oR and the surrounding environment are isotropic, lossless and nonmagnetic, with permittivities $\epsilon_{res}>\epsilon_{env}>0$ respectively. We use Gaussian CGS units and time dependence $e^{-i\omega t}$. Writing $k_0=\omega/c$, the source-free equations and dielectric interface conditions are
\begin{gather}
 \nabla\times\mathbf E=ik_0\mathbf H,\qquad
 \nabla\times\mathbf H=-ik_0\epsilon\mathbf E,\label{sup:mobius:eq:maxwell}\\
 \mathbf n_I\times[\mathbf E]=\mathbf n_I\times[\mathbf H]=0,\qquad
 \mathbf n_I\cdot[\epsilon\mathbf E]=\mathbf n_I\cdot[\mathbf H]=0.
 \label{sup:mobius:eq:interfaces}
\end{gather}
Here $c$ is the speed of light, $\mathbf n_I$ is the physical interface normal and brackets denote the jump across the boundary of the resonator. The transverse components form the column $F=(E_u,E_v,H_u,H_v)^T$. They coincide with the covariant components because $\mathbf g_u=\mathbf e_1$ and $\mathbf g_v=\mathbf e_2$. The longitudinal covariant components are
\begin{equation}
 E_s=fE_t-\Omega vE_u+\Omega uE_v,\qquad
 H_s=fH_t-\Omega vH_u+\Omega uH_v,
 \label{sup:mobius:eq:covariant-longitudinal}
\end{equation}
where $E_t=\mathbf E\cdot\mathbf t$ and $H_t=\mathbf H\cdot\mathbf t$. With the antisymmetric symbol normalized by $\varepsilon^{suv}=1$, the curl equations read
\begin{equation}
 \varepsilon^{ijk}\partial_jE_k=ik_0f g^{ij}H_j,\qquad
 \varepsilon^{ijk}\partial_jH_k=-ik_0\epsilon f g^{ij}E_j.
 \label{sup:mobius:eq:coordinate-maxwell}
\end{equation}
Their longitudinal components express $E_t,H_t$ through $F$,
\begin{equation}
 E_t[F]=\frac{i}{k_0\epsilon}(\partial_uH_v-\partial_vH_u),\qquad
 H_t[F]=\frac{1}{ik_0}(\partial_uE_v-\partial_vE_u).
 \label{sup:mobius:eq:reconstruction}
\end{equation}
For example, the transverse electric-curl equation gives
\begin{align}
 \partial_sE_u
 &=\partial_uE_s+ik_0(fH_v-\Omega uH_t)\nonumber\\
 &=\partial_u(fE_t)+ik_0fH_v
 +\Omega\big[(u\partial_v-v\partial_u)E_u+E_v\big].
 \label{sup:mobius:eq:curl-example}
\end{align}
Collecting all four transverse equations yields
\begin{equation}
 -i\partial_sF=(B_0+\Omega J+K(s))F,
 \label{sup:mobius:eq:material-propagation}
\end{equation}
where the straight-guide operator is
\begin{equation}
 B_0F=\begin{pmatrix}
 -i\partial_uE_t[F]+k_0H_v\\
 -i\partial_vE_t[F]-k_0H_u\\
 -i\partial_uH_t[F]-k_0\epsilon E_v\\
 -i\partial_vH_t[F]+k_0\epsilon E_u
 \end{pmatrix}.
 \label{sup:mobius:eq:straight-operator}
\end{equation}
The twist generator rotates both the profiles and their vector components~\cite{weiss2013},
\begin{equation}
 J=-i(u\partial_v-v\partial_u)I_4+
 \begin{pmatrix}\sigma_y&0\\0&\sigma_y\end{pmatrix},\qquad
 \sigma_y=\begin{pmatrix}0&-i\\i&0\end{pmatrix},
 \label{sup:mobius:eq:rotation-generator}
\end{equation}
with $I_4$ the four-component identity. The terms proportional to $\kappa$ are
\begin{align}
 K(s)&=\kappa(C_u\cos\vartheta-C_v\sin\vartheta),\nonumber\\
 C_uF&=-uB_0F+i(E_t[F],0,H_t[F],0)^T,\nonumber\\
 C_vF&=-vB_0F+i(0,E_t[F],0,H_t[F])^T.
 \label{sup:mobius:eq:curvature}
\end{align}
For example, $\partial_u(fE_t)=f\partial_uE_t-\kappa\cos\vartheta\,E_t$ gives the extra $E_t$ term in $C_uF$. Equations~\eqref{sup:mobius:eq:material-propagation}--\eqref{sup:mobius:eq:curvature} hold in each homogeneous region and are supplemented by Eq.~\eqref{sup:mobius:eq:interfaces}. At a wall $u=\pm a$, the continuous tangential components are $E_s,E_v,H_s,H_v$. At $v=\pm b$, they are $E_s,E_u,H_s,H_u$. Thus the interface conditions retain the geometry through Eq.~\eqref{sup:mobius:eq:covariant-longitudinal}. 

For the straight reference guide, $\Omega=\kappa=0$. Its separated modes and propagation eigenproblem are
\begin{equation}
 F(s,u,v)=e^{iq_\nu s}F_\nu(u,v),\qquad
 B_0F_\nu=q_\nu F_\nu.
 \label{sup:mobius:eq:straight-modes}
\end{equation}
The profiles satisfy the straight dielectric interface conditions and decay in the environment. At fixed $\omega$, the index $\nu$ labels a guided mode of the straight waveguide with transverse profile $F_\nu(u,v)$ and propagation constant $q_\nu$. For two guided modes, we define the power-flux form
\begin{equation}
 \langle F_\mu,F_\nu\rangle_P
 =\frac{c}{16\pi}\int_{\mathbb R^2}
 \bigl(\mathbf e_\mu^*\times\mathbf h_\nu
 +\mathbf e_\nu\times\mathbf h_\mu^*\bigr)
 \cdot\mathbf t\,du\,dv.
 \label{sup:mobius:eq:flux-form}
\end{equation}
Here $\mathbf e_\nu$ and $\mathbf h_\nu$ are the electric and magnetic profiles reconstructed from $F_\nu$. In the lossless straight guide, the fields decay in the environment, and the integrated cross flux is independent of $s$. The separated fields in Eq.~\eqref{sup:mobius:eq:straight-modes} give this flux a factor $e^{i(q_\nu-q_\mu)s}$, so $(q_\nu-q_\mu)\langle F_\mu,F_\nu\rangle_P=0$. Within each degenerate forward subspace we choose a flux-orthogonal basis. Normalizing each mode to unit power then gives $\langle F_\mu,F_\nu\rangle_P=\delta_{\mu\nu}$.

With material absorption, the guided power decreases along the waveguide and distinct modes need not be orthogonal under Eq.~\eqref{sup:mobius:eq:flux-form}. The finite ring can also radiate even when its permittivities are real. Its outgoing modes generally have complex resonance frequencies. We use Eq.~\eqref{sup:mobius:eq:flux-form} to normalize the straight-guide basis and assume negligible radiation over one circuit when calculating the leading ring and M\"obius resonance frequencies.

\subsection{Reflection symmetry and gallery indices}

The rectangle and its dielectric interface conditions are invariant under the transverse reflections. For $F=(E_u,E_v,H_u,H_v)^T$, their actions are
\begin{align}
 (I_uF)(u,v)&=\operatorname{diag}(-1,1,1,-1)F(-u,v),\nonumber\\
 (I_vF)(u,v)&=\operatorname{diag}(1,-1,-1,1)F(u,-v).
 \label{sup:mobius:eq:reflections}
\end{align}
These signs follow from the polar character of $\mathbf E$ and the axial character of $\mathbf H$. Both reflections commute with $B_0$ and with each other. Their product is a half-turn,
\begin{equation}
 \Pi=I_uI_v,\qquad (\Pi F)(u,v)=-F(-u,-v),\qquad \Pi^2=I.
 \label{sup:mobius:eq:half-turn}
\end{equation}
We choose modes with definite reflection characters,
\begin{equation}
 I_uF_\nu=m_{u,\nu}F_\nu,\qquad
 I_vF_\nu=m_{v,\nu}F_\nu,\qquad
 \Pi F_\nu=p_\nu F_\nu,\qquad
 p_\nu=m_{u,\nu}m_{v,\nu}=\pm1.
 \label{sup:mobius:eq:characters}
\end{equation}

We assign gallery indices from the nodal pattern of a transverse electric component at the operating dielectric contrast. After substituting Eq.~\eqref{sup:mobius:eq:reconstruction} into Eq.~\eqref{sup:mobius:eq:straight-operator}, the transverse operator and its straight-guide interface conditions are real. A mode with a simple real propagation eigenvalue can therefore be chosen with real transverse electric components.

Let $n_u,n_v$ count the sign-changing interior zeros of a nonzero component $e_\alpha$ on symmetric cuts across the resonator bulk in the $u$ and $v$ directions. The cuts have nonzero endpoint values and do not lie on a nodal line. Zeros away from the center occur in reflection-related pairs. An odd component has one additional sign change at the center, so its scalar reflection signs are $(-1)^{n_u}$ and $(-1)^{n_v}$. Since the half-turn also reverses the transverse vector component,
\begin{equation}
 p_\nu=-(-1)^{n_u+n_v}=(-1)^{n_u+n_v+1}.
 \label{sup:mobius:eq:nodal-parity}
\end{equation}
Using the selected electric component, we assign the gallery indices as
\begin{equation}
 (r,z)=
 \begin{cases}
 (n_u+1,n_v),&e_\alpha=e_u,\\
 (n_u,n_v+1),&e_\alpha=e_v.
 \end{cases}
 \qquad p_\nu=(-1)^{r+z}.
 \label{sup:mobius:eq:gallery-indices}
\end{equation}
Thus the no-node $u$- and $v$-polarized patterns carry labels $(1,0)$ and $(0,1)$ and have $p=-1$. A $u$-polarized pattern with one sign-changing $u$-node and no $v$-node has label $(2,0)$ and $p=+1$. These indices describe nodal families. Different modes can share the same $(r,z)$ and remain distinguished by $\nu$. Changing a generic cut can change the nodal count only by an even number, leaving Eq.~\eqref{sup:mobius:eq:gallery-indices} unchanged modulo two. The parity relation uses the full dielectric fields and requires no weak-guidance approximation.

\subsection{Adiabatic transport and geometric phase}

Each reflection reverses the sense of rotation, while a half-turn reverses the curvature direction. Equations~\eqref{sup:mobius:eq:rotation-generator} and \eqref{sup:mobius:eq:curvature} therefore give
\begin{equation}
 I_iJI_i^{-1}=-J\quad(i=u,v),\qquad
 \Pi C_u\Pi^{-1}=-C_u,\qquad
 \Pi C_v\Pi^{-1}=-C_v.
 \label{sup:mobius:eq:geometric-parities}
\end{equation}
The flux form is invariant under these transformations. Applying them to a mode with the characters in Eq.~\eqref{sup:mobius:eq:characters} gives
\begin{equation}
 \langle F_\nu,JF_\nu\rangle_P=0,\qquad
 \langle F_\nu,C_uF_\nu\rangle_P=
 \langle F_\nu,C_vF_\nu\rangle_P=0.
 \label{sup:mobius:eq:zero-diagonal}
\end{equation}
The same covariance holds for the complete interface problem. At zero curvature, a transverse reflection maps the twist $\Omega$ to $-\Omega$, including the tangential traces in Eq.~\eqref{sup:mobius:eq:covariant-longitudinal}. At fixed twist, $\Pi$ reverses the transverse curvature vector. For an isolated mode, these symmetries make the first-order shift of $q_\nu$ vanish, including the contribution from the dielectric interfaces. The off-diagonal matrix elements of $\Omega J+K(s)$ can still mix modes at first order. At a coupled degeneracy they can also split the propagation constants at first order.

We take the transverse extent of the guided field, including its environmental tail, to be small compared with $R$. Near the square cross section, Fig.~3 of the main text shows closely spaced pairs of resonances with longitudinal indices $m=n+\frac{1}{2}$ and $m=n+\frac{3}{2}$, where $n\in \mathbb{Z}$. We consider the branches away from this interaction. As $R$ increases at fixed $N$ and cross section, we assume that $q_\nu$ stays separated from the radiation threshold $k_0\sqrt{\epsilon_s}$ and that $|q_\nu-q_\mu|$ has a positive lower bound independent of $R$ for every mode $\mu\ne\nu$ coupled to $\nu$.

Since $\Omega=N/(2R)$ and $\kappa=1/R$, the matrix elements $V_{\mu\nu}=\langle F_\mu,(\Omega J+K(s))F_\nu\rangle_P$ are $O(R^{-1})$. Equation~\eqref{sup:mobius:eq:zero-diagonal} gives $V_{\nu\nu}=0$. The first-order admixture of mode $\mu$ is proportional to $V_{\mu\nu}/(q_\nu-q_\mu)$. Coupling this admixture back to mode $\nu$ gives a contribution to its propagation constant proportional to $V_{\nu\mu}V_{\mu\nu}/(q_\nu-q_\mu)$, which is $O(R^{-2})$. Integration over $L=2\pi R$ gives
\begin{equation}
 \Delta q_\nu=O(R^{-2}),\qquad
 \int_0^L\Delta q_\nu(s)\,ds=O(R^{-1}),\qquad
 \Omega L=N\pi.
 \label{sup:mobius:eq:asymptotic-control}
\end{equation}
Here $\Delta q_\nu$ is the correction to the local propagation constant. Its contribution to the phase over one circuit vanishes as $R$ increases, while the total rotation of the cross section remains $N\pi$.

To calculate the corresponding geometric phase, we represent the transverse profiles in the periodic Frenet frame. If $(x,y)$ are coordinates along $\mathbf n,\mathbf b$, then $(x,y)^T=Q_\vartheta(u,v)^T$, where
\begin{align}
 Q_\vartheta&=\begin{pmatrix}\cos\vartheta&-\sin\vartheta\\
 \sin\vartheta&\cos\vartheta\end{pmatrix},\qquad
 \mathcal R(\vartheta)=e^{-i\vartheta J},\nonumber\\
 [\mathcal R(\vartheta)F](x,y)
 &=\operatorname{diag}(Q_\vartheta,Q_\vartheta)
 F\bigl(Q_{-\vartheta}(x,y)^T\bigr).
 \label{sup:mobius:eq:active-rotation}
\end{align}
The rotation carries the dielectric profile and its interface conditions with the field. In the leading local straight-guide problem,
\begin{equation}
 B_{\rm loc}(s)=\mathcal R(\Omega s)B_0\mathcal R^{-1}(\Omega s),\qquad
 f_\nu(s)=\mathcal R(\Omega s)F_\nu,\qquad
 B_{\rm loc}(s)f_\nu(s)=q_\nu f_\nu(s).
 \label{sup:mobius:eq:local-operator}
\end{equation}
Writing the Frenet-frame field as $\Psi=\mathcal R(\Omega s)F$ in Eq.~\eqref{sup:mobius:eq:material-propagation} cancels the explicit $\Omega J$ term against the derivative of $\mathcal R$. In Eq.~\eqref{sup:mobius:eq:local-operator}, the local mode satisfies the dielectric interface conditions of a straight guide with the cross section held at its orientation at $s$. For example, on the actual wall $u=\pm a$, continuity of $E_s$ and $E_v$ gives $[E_t]=(\Omega v/F)[E_u]$ by Eq.~\eqref{sup:mobius:eq:covariant-longitudinal}. The straight-guide condition $[E_t]=0$ is the leading approximation for $|\Omega v|/F\ll1$. Using this local profile and neglecting the curvature correction at the order specified in Eq.~\eqref{sup:mobius:eq:asymptotic-control}, we expand the field in forward guided modes,
\begin{equation}
 -i\partial_s\Psi=B_{\rm loc}(s)\Psi,\qquad
 \Psi(s)=\sum_\nu a_\nu(s)e^{iq_\nu s}f_\nu(s).
 \label{sup:mobius:eq:local-propagation}
\end{equation}
Substitution and projection with Eq.~\eqref{sup:mobius:eq:flux-form} yield
\begin{align}
 \frac{d}{ds}a_\mu&=-\sum_\nu a_\nu e^{i(q_\nu-q_\mu)s}
 \langle f_\mu,\partial_sf_\nu\rangle_P,\nonumber\\
 \langle f_\mu,\partial_sf_\nu\rangle_P
 &=-i\Omega J_{\mu\nu},\qquad
 J_{\mu\nu}=\langle F_\mu,JF_\nu\rangle_P.
 \label{sup:mobius:eq:amplitude-transport}
\end{align}
Integration of an off-diagonal term in Eq.~\eqref{sup:mobius:eq:amplitude-transport} gives a modal amplitude of order $|\Omega J_{\mu\nu}/(q_\nu-q_\mu)|$. The curvature matrix elements are $K_{\mu\nu}(s)=\langle F_\mu,K(s)F_\nu\rangle_P$, with $K(s)$ given by Eq.~\eqref{sup:mobius:eq:curvature}. For each coupled pair, the sufficient local conditions are
\begin{equation}
 |\Omega J_{\mu\nu}|\ll|q_\nu-q_\mu|,\qquad
 \max_s|K_{\mu\nu}(s)|\ll|q_\nu-q_\mu|,\qquad \mu\ne\nu.
 \label{sup:mobius:eq:adiabatic-conditions}
\end{equation}
The accumulated phase corrections must also be small as in Eq.~\eqref{sup:mobius:eq:asymptotic-control}. For the fundamental modes $u$ and $v$, Eq.~\eqref{sup:mobius:eq:adiabatic-conditions} requires the twist coupling $|\Omega J_{uv}|$ to be small compared with the separation $|q_u-q_v|$ of their straight-guide propagation constants. When the splitting becomes comparable to the coupling near the square cross section, both polarizations must be retained.

For an isolated mode in the adiabatic limit, Eq.~\eqref{sup:mobius:eq:zero-diagonal} makes the diagonal connection vanish and Eq.~\eqref{sup:mobius:eq:amplitude-transport} gives $\frac{d}{ds}a_\nu=0$ at this order. Since $\mathcal R(N\pi)=\Pi^N$, its endpoint is $f_\nu(L)=p_\nu^Nf_\nu(0)$. The geometric phase is~\cite{berry1984},
\begin{align}
 \gamma_\nu^{\rm B}
 &=\arg\langle f_\nu(0),f_\nu(L)\rangle_P
 +i\int_0^L\langle f_\nu,\partial_sf_\nu\rangle_P\,ds\nonumber\\
 &=N\pi\frac{1-p_\nu}{2}\pmod{2\pi}.
 \label{sup:mobius:eq:berry-parity}
\end{align}
A change $f_\nu(s)\mapsto e^{i\chi(s)}f_\nu(s)$ adds $\chi(L)-\chi(0)$ to the endpoint phase and subtracts the same quantity from the integral. With the nodal indices of Eq.~\eqref{sup:mobius:eq:gallery-indices},
\begin{equation}
 \gamma_{rz}^{\rm B}=N\pi(r+z)\pmod{2\pi}.
 \label{sup:mobius:eq:berry-indices}
\end{equation}
Odd $N(r+z)$ gives phase $\pi$, while even $N(r+z)$ gives zero. This is the adiabatic phase of the isolated transverse family.

\subsection{Resonance closure and comparison with the ring}

The Frenet frame returns to itself after one circuit. Single-valuedness of the leading field $e^{iq_\nu s}f_\nu(s)$ therefore requires
\begin{equation}
 e^{iq_\nu L}f_\nu(L)=f_\nu(0)
 \quad\Longrightarrow\quad
 e^{iq_\nu L}p_\nu^N=1
 \quad\Longrightarrow\quad
 q_\nu L+\gamma_\nu^{\rm B}=2\pi n,\qquad n\in\mathbb Z.
 \label{sup:mobius:eq:closure}
\end{equation}
The allowed longitudinal indices $m=q_\nu R$ belong to the lattice
\begin{equation}
 m\in\mathbb Z+\frac{N(r+z)}2.
 \label{sup:mobius:eq:longitudinal-indices}
\end{equation}
The sign of the half-integer displacement does not change this lattice. In an untwisted ring with the same radius, cross section and dielectric constants, the leading propagation constant is the same $q_\nu$ and the closure condition is $q_\nu L=2\pi\ell$. The curvature corrections vanish in the large-$R$ limit described by Eq.~\eqref{sup:mobius:eq:asymptotic-control}.

For the forward guided mode chosen in Eq.~\eqref{sup:mobius:eq:flux-form}, the group velocity is $v_{g,\nu}=P_\nu/U_\nu$, where $P_\nu>0$ is the power carried along the guide and $U_\nu>0$ is the energy per unit length. Hence $\partial_\omega q_\nu=1/v_{g,\nu}>0$, and the ring resonance frequency increases with longitudinal index $\ell$. For odd $N(r+z)$, the M\"obius resonance between two adjacent ring resonances of the same family obeys
\begin{equation}
 q_\nu\bigl(\omega_{R,\nu}^{(\ell)}\bigr)=\frac\ell R,\qquad
 q_\nu(\omega_{M,\nu})=\frac{\ell+1/2}{R},\qquad
 q_\nu\bigl(\omega_{R,\nu}^{(\ell+1)}\bigr)=\frac{\ell+1}{R}.
 \label{sup:mobius:eq:half-step}
\end{equation}
The free spectral range, denoted by $\Xi$, is the angular-frequency spacing between these ring resonances,
\begin{equation}
 \Xi=\omega_{R,\nu}^{(\ell+1)}-\omega_{R,\nu}^{(\ell)}.
 \label{sup:mobius:eq:free-spectral-range}
\end{equation}
Between adjacent ring resonances the propagation phase $q_\nu L$ increases by $2\pi$. At the intervening M\"obius resonance it has increased by $\pi$ from its value at the lower ring resonance Eq.~\eqref{sup:mobius:eq:half-step}. If the slope $\partial_\omega q_\nu$ varies little across this interval, the full phase increase satisfies $L(\partial_\omega q_\nu)\Xi\simeq2\pi$. The increase of $\pi$ then takes approximately half the frequency interval. With $L=2\pi R$, we obtain
\begin{equation}
 \Xi\simeq\frac{1}{R\,\partial_\omega q_\nu},\qquad
 \omega_{M,\nu}-\omega_{R,\nu}^{(\ell)}\simeq\frac F2,\qquad
 \omega_{M,\nu}\simeq
 \frac{\omega_{R,\nu}^{(\ell)}+\omega_{R,\nu}^{(\ell+1)}}2.
 \label{sup:mobius:eq:half-fsr}
\end{equation}
The half-step in $q_\nu R$ follows from the geometric phase. To convert it to a frequency shift, we used the slope $\partial_\omega q_\nu$ at the lower ring resonance $\omega_{R,\nu}^{(\ell)}$. Across the spacing $\Xi$, the slope changes by at most $\Xi\max|\partial_\omega^2q_\nu|$, where the maximum is taken between the two ring resonances. The linear approximation in Eq.~\eqref{sup:mobius:eq:half-fsr} requires this change to be small compared with the slope at the lower resonance. Using the expression for $\Xi$ gives
\begin{equation}
    \frac{\Xi\max|\partial_\omega^2q_\nu|}{\partial_\omega q_\nu}
    \simeq
    \frac{\max|\partial_\omega^2q_\nu|}
    {R(\partial_\omega q_\nu)^2}
    \ll 1.
\end{equation}
Comparing with the upper ring resonance instead gives $-\Xi/2$. For even $N(r+z)$, the two resonators have the same integer longitudinal lattice at leading order. The fundamental families $(1,0)$ and $(0,1)$ are both shifted by $\Xi/2$ for odd $N$, while $(2,0)$ has no leading geometric shift.